\documentclass[11pt,a4paper]{article}
\pdfoutput=1
\usepackage{jheppub, hyperref, cleveref}
\usepackage{amsfonts, amsmath, amssymb, latexsym}
\usepackage{graphicx, caption, subcaption}
\usepackage[normalem]{ulem}
\usepackage[dvipsnames]{xcolor}

\title{Improved reconstruction of a stochastic gravitational wave background with LISA}

\author[a]{ Raphael Flauger}
\author[b]{\!\!, Nikolaos Karnesis}
\author[c]{\!\!, Germano Nardini}
\author[d]{\!\!, Mauro Pieroni\footnote{Project coordinator and corresponding author: m.pieroni@imperial.ac.uk}}
\author[e,f]{\!\!, Angelo Ricciardone}
\author[g]{\!\!, Jes\'us Torrado}

\author[]{\\ \centering \texttt{(For the LISA Cosmology Working Group)}}

\affiliation[a]{UC  San  Diego,  Department  of  Physics,  9500  Gilman  Rd,  La  Jolla,  CA,  92093,  USA}

\affiliation[b]{APC, Universit\'e Paris Diderot, CNRS/IN2P3, CEA/Irfu, Observatoire de Paris, Sorbonne Paris Cit\'e, 10 rue Alice Domont et L\'eonie Duquet, 75013 Paris, France}

\affiliation[c]{Department of Mathematics and Physics, University of Stavanger, NO-4036 Stavanger, Norway}

\affiliation[d]{Blackett Laboratory, Imperial College London, SW7 2AZ, UK}

\affiliation[e]{Dipartimento di Fisica e Astronomia ``G. Galilei",
Universit\`a degli Studi di Padova, via Marzolo 8, I-35131 Padova, Italy}

\affiliation[f]{INFN, Sezione di Padova, via Marzolo 8, I-35131 Padova, Italy}

\affiliation[g]{Institute for Theoretical Particle Physics and Cosmology (TTK), RWTH Aachen University, D-52056 Aachen, Germany}

\abstract{
We present a data analysis methodology for a model-independent reconstruction of the spectral shape of a stochastic gravitational wave background with LISA. We improve a previously proposed reconstruction algorithm that relied on a single Time-Delay-Interferometry (TDI) channel by including a complete set of TDI channels. As in the earlier work, we assume an idealized equilateral configuration.
We test the improved algorithm with a number of case studies, including reconstruction in the presence of two different astrophysical foreground signals. We find that including additional channels helps in different ways: it reduces the uncertainties on the reconstruction; it makes the global likelihood maximization less prone to falling into local extrema; and it efficiently breaks degeneracies between the signal and the instrumental noise. }

\begin{document}
\begin{figure}
\begin{flushright}
\href{https://lisa.pages.in2p3.fr/consortium-userguide/wg_cosmo.html}{\includegraphics[width = 0.2 \textwidth]{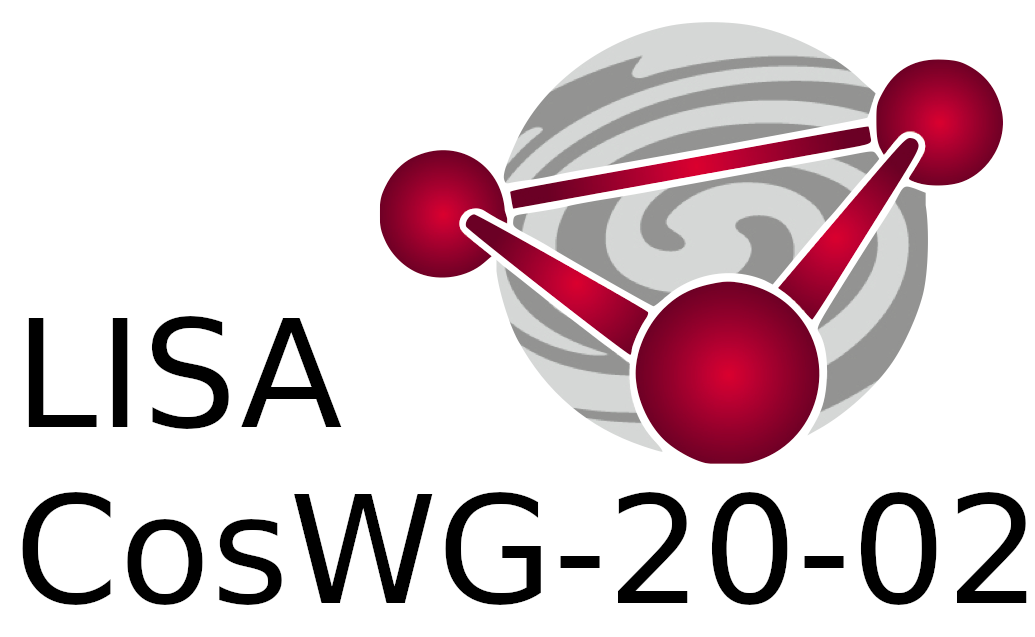}}\\[5mm]
TTK-20-29
\end{flushright}
\end{figure}

\maketitle

\section{ Introduction}
\label{sec:intro}
The {\it Laser Interferometer Space Antenna} (LISA) experiment~\cite{Audley:2017drz}, a European Space Agency (ESA) mission with National Aeronautics and Space Administration (NASA) partnership, will be the milestone for Gravitational Wave (GW) detection in space. It is planned to launch in the mid-2030s and it will consist of a constellation of three satellites forming a nearly equilateral triangle with 2.5 million km length arms. By monitoring the relative displacements among the three satellites, LISA will perform three correlated interferometry measurements (typically dubbed XYZ channels), which can be transformed into three uncorrelated data streams (typically dubbed A, E, and T). LISA will probe GWs in the milli-Hertz regime which is not accessible with present and future ground-based detectors~\cite{TheLIGOScientific:2014jea, Advanced-Virgo, Hild:2010id, Somiya:2011np, Aso:2013eba, Evans:2016mbw}, and will open a new window for GW physics.

The sky is extremely rich in the LISA sensitivity band, with order of ten thousand sources that will be individually resolved during the (at least) four-year mission duration~\cite{Audley:2017drz}. Depending on the astrophysical formation scenarios~\cite{Sesana:2004sp, Sesana:2004gf, Barausse:2012fy, Klein:2015hvg, Bonetti:2018tpf}, LISA is going to detect few hundreds of massive ($\sim 10^4$--$10^7 M_\odot$) black hole binary mergers with Signal-to-Noise Ratio (SNR) up to thousand. These likely constitute the strongest sources that LISA will resolve individually. Besides them, LISA will resolve between a few to a few hundred extreme mass ratio inspirals per year~\cite{Babak:2017tow} and up to tens of thousands of galactic white dwarf binaries~\cite{Nelemans:2003ha, Ruiter:2007xx, Marsh:2011yj, Korol:2017qcx, Korol:2018wep}. 

The existence of so many resolvable sources is accompanied by the presence of a large number of events that will not be resolved individually, leading to the generation of a Stochastic GW Background (SGWB). Given the LISA detection performances, there are two guaranteed astrophysical background components in the LISA band: the component sourced by the unresolved compact Galactic binaries, overcoming the instrumental noise at the frequencies $2 \times 10^{-4} - 2 \times 10^{-3}$\, Hz~\cite{Bender:1997hs, Evans:1987qa}, and the component due to neutron star and stellar-origin black hole mergers~\cite{LIGOScientific:2019vic}, relevant in the range $2 \times 10^{-3} - 1 \times 10^{-2}$\, Hz. Beyond these, it is plausible to expect an extra component originated by the extreme mass ratio inspirals, possibly above the LISA sensitivity at about $10^{-3} -  10^{-2}$\, Hz~\cite{Bonetti:2020jku}.

On top of these astrophysical sources, LISA will potentially be sensitive to SGWBs related to physics of the early universe or  physics of particles beyond the standard model~\cite{DallArmi:2020dar}. (For recent reviews of non-astrophysical sources of SGWB, see e.g.~refs.~\cite{Caprini:2018mtu, Barausse:2020rsu}). Commonly considered examples include superradiance effects~\cite{Barausse:2018vdb, Alexander:2018qzg}, cosmological first order phase transitions~\cite{Caprini:2015zlo, Caprini:2019egz}, networks of topological defects (e.g. cosmic strings)~\cite{Auclair:2019wcv} and even inflationary models~\cite{Bartolo:2016ami} with and without subsequent SGWB due to primordial black holes~\cite{Garcia-Bellido:2016dkw}. In practice, the actual SGWB signal may be a superposition of any of these sources --- and possibly of previously unknown ones --- with the SGWB components of astrophysical origin.

In order to extract as much information about the processes sourcing the SGWB as possible, a detailed signal characterization is required. This relies on an accurate reconstruction of the overall SGWB signal and the capability of precisely breaking it into its components. The Galactic component is expected to be anisotropic and to have a yearly modulation which helps in (partially) 
disentangling it from the isotropic components~\cite{Adams:2013qma}.\footnote{The angle between the Galactic disk and the plane defined by the three LISA's satellites varies while LISA orbits around the Sun. This is at origin of the modulation.} The astrophysical extra-Galactic component is instead expected to be almost isotropic and stationary, so that separating it from other isotropic and stationary signals is challenging. However, it has recently been shown that it is possible to employ techniques to perform an accurate separation of this component from other SGWBs~\cite{Pieroni:2020rob}. It is worth stressing that caution is necessary when moving towards the treatment of realistic LISA SGWB data due to the complexity of the instrumental noise,  the way LISA measures the signals, and the possibility that an unexpected, unmodeled SGWB contaminates the data. For example, we should expect instrumental noise non-stationarities such as noise transients~\cite{PhysRevD.99.024019, PhysRevLett.120.061101} and spectral lines (as seen already in LISA Pathfinder data~\cite{PhysRevLett.116.231101}), and slow variations of the noise power spectrum due to thermal gradients inside the spacecrafts~\cite{Gibert_2015, 10.1093/mnras/stz1017}. In addition, non-stationarities and/or non-Gaussianity can arise from the non-perfect subtraction of bright GW transient signals~\cite{Ginat:2019aed}. There are various strategies to mitigate, or take into account these effects, such as adopting higher-tail likelihood functions, or fitting a more sophisticated noise power spectrum model. In this work, we shall assume a simplified scenario, where the instrumental noise is Gaussian and stationary, and that we are working with perfect residuals, which means that bright sources have been subtracted perfectly from the data-stream.

In the same spirit as ref.~\cite{Karnesis:2019mph}, and following up on ref.~\cite{Caprini:2019pxz}, we employ a model-independent approach to reconstruct the frequency shape of an unknown SGWB. Our procedure is based on and extends the \texttt{SGWBinner} code~\cite{Caprini:2019pxz}, which divides the frequency range of LISA into sub-intervals (\textit{bins}) in which we the signal is approximated by a power law.\footnote{Notice that in log-log scale this is the equivalent to performing a series of Taylor expansions truncated at linear order.} By starting with an arbitrary number of initial bins and subsequently combining them according to information criteria to avoid over-fitting, our procedure is robustly able to capture the frequency dependence of the spectrum, and leads to an accurate reconstruction of the injected signal~\cite{Caprini:2019pxz}. The analysis carried out in ref.~\cite{Caprini:2019pxz} relied on a single data stream (in practice ref.~\cite{Caprini:2019pxz} used the X channel). In this work we consider the full set of correlated channels X, Y, and Z. As usual, we perform the diagonalization of these data streams and describe the signal and noise within the data in the so-called AET basis~\cite{Hogan:2001jn, Adams:2010vc}. As we show in this work (where some simplified assumption are made), this allows us to break degeneracies between the signal and noise spectra which helps in disentangling the two components.\footnote{ In this analysis we work assuming that the detector arm lengths are equal. This allows to state that in the AET basis the noise is orthogonal if identical power spectral density in every noise component (e.g.~the noise reduction system and phasemeters) on each spacecraft are considered. The case where the noise is not orthogonal is described in ref.~\cite{Adams:2010vc}.}

We test the model independent reconstruction, using the AET channels, on three well motivated SGWB signals: a simple power law that can mimic an astrophysical GW signal coming from numerous unresolved binary mergers~\cite{LIGOScientific:2019vic, Sesana:2016ljz} and/or a cosmological signal coming from some inflationary mechanisms~\cite{Bartolo:2016ami} or cosmic strings~\cite{Auclair:2019wcv}; a broken power-law signal approximating the SGWB from a first order phase transition in the early Universe~\cite{Caprini:2015zlo, Caprini:2019egz}; and finally a bump signal, which is expected mostly from post-inflationary sources of GWs, like preheating~\cite{Bartolo:2016ami}. In the present paper our primary focus is the development of a model-independent method to reconstruct the overall homogeneous and isotropic SGWB signal, without aspiring to address the issue of the component separation. Nevertheless, in order to prepare a method that remains robust even when backgrounds and foregrounds coexist, we perform some tests on the method's reconstruction capabilities also in the presence of some illustrative foregrounds. Further improvements which would make the analysis more realistic and could degrade our reconstructions are mentioned.

This paper is structured as follows. In~\cref{sec:detection} we present our formalism for modeling the LISA data stream in terms of a signal and a noise component, and we introduce our mock signals. In~\cref{sec:data_analysis} we explain our data analysis techniques and outline the main algorithm of the \texttt{SGWBinner} code introduced in ref.~\cite{Caprini:2019pxz}. In~\cref{sec:results} we present the results obtained by applying our techniques to a set of mock signals and in~\cref{sec:foreground} we test the robustness of the results in the presence of foregrounds. Finally in~\cref{sec:conclusions} we draw our conclusions. Some technical details are provided in~\cref{sec:technicalities}.

\section{SGWB measurements with LISA}
\label{sec:detection}

The main observable of interest for the detection and characterization of an isotropic SGWB is its power spectrum $P_h(k)$. In this section we provide a brief summary on how LISA measures the SGWB power spectrum.

\subsection{Spectral densities and power spectra}
\label{sec:power_spectra}
We begin with the assumption that we are working with `perfect' residuals, which means that all transient signals and glitches in the noise have been subtracted from the time stream. After this procedure, the time stream $d(t)$ contains noise, which we denote by $n(t)$, and
a residual stochastic signal $s(t)$. We assume that both the noise and the residual signal are stationary. Because of periodic antenna re-pointing and other operational interruptions, the data are expected to be broken up into segments of length $T$. For the sake of comparison with the previous study~\cite{Caprini:2019pxz}, we take $T=11.5$ days.\footnote{The value of $T$ is still under discussion, and in practice not all segments may have the same duration. In the stationary regime, the duration $T$ of the segments is fictitious. We however prefer to keep this format to prepare the formalism and the code for future developments where the stationary assumption between a segment and the subsequent one is broken.} 

In practice the data stream $d(t)=s(t)+n(t)$ is sampled at a finite rate, but we model it as a real-valued function on the interval $\left[ -T/2, T/2\right]$ to keep the notation more appealing. 

We assume that signal and noise are uncorrelated. We can then treat them separately and similarly. First, we define the Fourier transform
\begin{equation}
\tilde{s}\left(f\right) =  \int_{-T/2}^{T/2} d t  \;\textrm{e}^{ 2\pi i f t} s \left(t\right)\,.
\end{equation}
For stationary $s(t)$, the ensemble average of Fourier modes must obey
\begin{equation}
\label{eq:singlesidedsps}
\hspace{-1cm}	\langle \tilde{s}(f) \tilde{s}^*(f') \rangle \equiv \frac{1}{2} \delta \left(f-f'\right) S\left(f\right) \; , 
\end{equation}
for real and positive $S\left(f\right)$. Furthermore, since $s(t)$ is real, the Fourier transform obeys the reality condition $\tilde{s}^*\left(f\right)=\tilde{s}\left(-f\right)$, so that $S(f)=S(-f)$. For this reason, it is conventional to work with the ``one-side'' power spectrum $S(f)$ defined only for positive (physical) frequencies, and to include a factor $1/2$ as we have done here.

So far we have considered a single channel, that is, the Time Delay Interferometry (TDI)  measurement exploiting only two arms of LISA. However, with its three arms LISA will provide three TDI channels, called $\{ \text{X,Y,Z}\}$ or $\{\text{A,E,T}\}$, depending on the basis adopted (see sec.~\ref{sec:three_channels} for the relationship between the two bases). Equation \eqref{eq:singlesidedsps} then generalizes to 
\begin{equation}
\hspace{-1cm}	\langle \tilde{s}_i(f) \tilde{s}_j^*(f') \rangle \equiv \frac{1}{2} \delta \left(f-f'\right) S_{ij}\left(f\right) \; , 
\end{equation}
where the latin indices run over the basis $\{\text{X,Y,Z}\}$ or $\{\text{A,E,T}\}$, and the spectral densities $S_{ij}\left(f\right)$ are characterized by a Hermitian matrix-valued function that is related to the power spectrum $P_h(f)$ through the respective response functions $\mathcal{R}_{ij}$: 
\begin{equation}
		\label{eq:two_signal_correlators}
		S_{ij}\left(f\right) =  \mathcal{R}_{ij} (f) P_h (f) \; .
\end{equation}
The response functions for the different channel cross-spectra are shown in~\cref{sec:noise_and_signal_model}, and their derivation is provided in~\cref{sec:technicalities}. Notice that \cref{eq:two_signal_correlators} quotes the relationship in the limit of negligible relative motion of the LISA's spacecrafts. For our purpose, the matrix $\mathcal{R}_{ij} (f)$ is real and symmetric (see e.g.~ref.~\cite{Domcke:2019zls} for some peculiar scenarios where this property does not occur).

Likewise, for stationary (and real) noise $n(t)$ one concludes
\begin{equation}
\label{eq:singlesidednps}
\hspace{-1cm}	\langle \tilde{n}_i (f) \tilde{n}_j^*(f') \rangle = \frac{1}{2} \delta \left(f-f'\right) N_{ij}\left(f\right) \;,
\end{equation}
where $N_{ij}(f)$ is a Hermitian matrix function of {\it single-sided noise power spectra}.\footnote{In the literature the functions $N_{ij}(f)$ are often denoted by $P_{n}$. Their dimension is $\rm{Hz}^{-1}$ since both $\tilde{n}(f)$ and $\delta$ have dimension $\rm{Hz}^{-1}$.}

For the SGWB of primordial origin, it is common practice to predict the signal in terms of $\Omega_{\rm GW}(f)$, the energy density per logarithmic frequency interval scaled by the critical density.  The noise spectrum in $\Omega_{\rm GW}(f)$ is related to the noise spectrum in~\cref{eq:singlesidednps} by
\begin{equation}
		N_{ij}^{\Omega} (f) = \frac{4 \pi^2 f^3}{3 (H_0/h)^2} N_{ij} (f) \;,
\end{equation}
with $H_0$ denoting today's Hubble parameter whose dimensionless parameter normalization is $h$. The power spectrum of the signal can be recast in units of $\Omega_{\rm GW}(f)$ in the same way.

\subsection{LISA noise model}
\label{sec:noise_and_signal_model}
So far our discussion of signal and noise power spectral densities generally applies to any gravitational wave interferometer. We now specialize our discussion to LISA, beginning with the noise model. The TDI variables are designed to eliminate the dominant source of noise caused by fluctuations in the central frequency of the laser as well as the noise caused by displacements of the optical benches. In this simplified model, the residual noise components that enter into each TDI channel can be grouped into two effective quantities, dubbed ``Interferometry Metrology System" (IMS) noise, which, for example, includes shot noise, and ``acceleration" noise associated with the random displacements of the proof masses caused, for example, by local environmental disturbances. Our current understanding of the LISA noise is based on the LISA Pathfinder experiment~\cite{PhysRevLett.116.231101} and laboratory tests. The IMS and acceleration noise power spectra are given by~\cite{ldcdoc}
\begin{equation}
	\begin{aligned}
	P_{\rm IMS}(f, P) &= P^2~ \frac{{\rm pm}^2}{\rm Hz}\left[1 + \left(\frac{2\,\textrm{mHz}}{f} \right)^4  \right] \left(\frac{2 \pi f}{c} \right)^2\; , \\
	P_{\rm acc}(f, A) &= A^2~ \frac{{\rm fm}^2}{{\rm s}^4\,{\rm Hz}}\left[1 + \left(\frac{0.4\,\textrm{mHz}}{f} \right)^2  \right] \left[1 + \left(\frac{f}{8\,\textrm{mHz}} \right)^4  \right] \left(\frac{1}{2 \pi f } \right)^4 \left(\frac{2 \pi f}{c} \right)^2 \; .
	\end{aligned}
\label{eq:acc_int_noise}
\end{equation}
\begin{figure}[thb]
    \centering
    \includegraphics[width = 0.47 \textwidth]{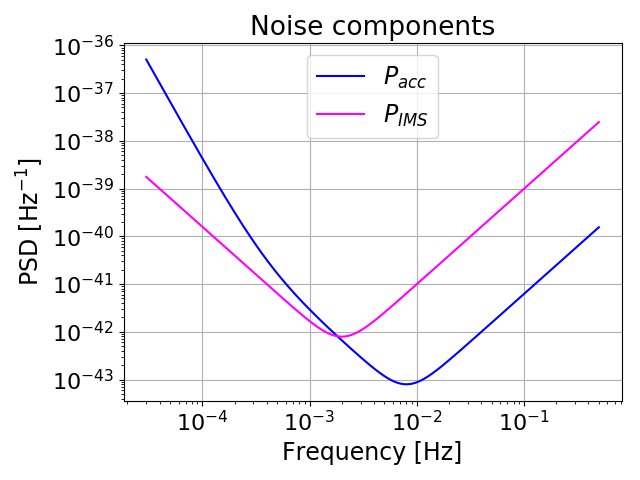}
    \includegraphics[width = 0.47 \textwidth]{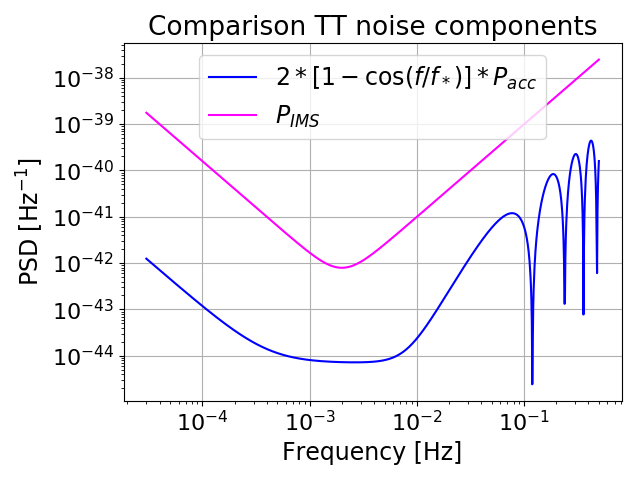}
    \caption{Left panel: the IMS and acceleration noise power spectra expressed in \cref{eq:acc_int_noise}. Right panel: The IMS and acceleration contributions to $N_{\text{TT}}$ weighted with their pre-factors as in \cref{eq:psdTT}. In both panels we fix $P=15$ and $A=3$.}
    \label{fig:PSD_comparisons}
\end{figure}
\noindent ESA's mission specifications require the amplitudes to be $P=15$ and $A= 3$ with $\pm $20\% margins. The two spectra are shown in the left panel of~\cref{fig:PSD_comparisons} for the central values of $P$ and $A$. 

The precise determination of the noise properties is one of the main technical challenges of the LISA mission, and it seems premature to attempt to keep track of the full complexity here. For the IMS contribution, we make the simplifying assumption that the noise spectra for all links are identical, stationary, and uncorrelated. Similarly, for the acceleration noise,  we assume that the fluctuations of  the  masses  are  isotropic and stationary,  that  the  power spectra  for all test  masses are equal, and  that  the fluctuations  of  the  different  masses are uncorrelated. Furthermore, we assume that the three satellites form an equilateral triangle, $L_1=L_2=L_3=L=2.5 \times 10^9\,{\rm m}$.  
Under these assumptions, the total power spectral density for the noise auto-correlation is
\begin{equation}
	N_{aa}(f, A, P) = 16 \sin^2\left(\frac{2 \pi f L}{c}\right) \left\{  \left[3 +\cos \left(\frac{4 \pi f L}{c} \right)\right] P_{\rm acc}(f, A) + P_{\rm IMS}(f, P) \right\} \; ,
	\label{eq:N_XX}
\end{equation}
and the noise cross-spectra are
\begin{equation}
	N_{ ab }(f, A, P) = -8 \sin^2\left(\frac{2 \pi f L}{c}\right) \cos \left(\frac{2 \pi f L}{c}\right)  \left[4 P_{\rm acc}(f, A) +  P_{\rm IMS}(f, P) \right] \; ,
	\label{eq:N_XY}
\end{equation}
where $a,b\in\{$X,Y,Z$\}$ and $a\ne b$. Notice, in particular, that for our assumptions, the noise covariance matrix is real. For completeness, we include a derivation in~\cref{sec:noise}. 

As we explain below, we marginalize over the amplitude of the IMS and acceleration noise power spectra in our analysis. However, we take the functional form of the noise model used to generate the mock data to be the same as the model used in fitting the data. So differences between the functional form of the instrumental noise and the noise model would introduce a bias and will have to be closely monitored. This is particularly true for LISA, for which calibration of the noise must happen at the same time as the reconstruction of the SGWB and other signals.

\subsection{AET basis}
\label{sec:three_channels}

So far we have defined the signal and noise spectra in~\cref{sec:power_spectra} in an arbitrary basis, and have introduced the LISA noise spectra in the XYZ basis in~\cref{sec:noise_and_signal_model}. We now introduce another commonly used basis of TDI channels, the AET basis, which diagonalizes the signal and noise covariance matrices~\cite{Hogan:2001jn, Adams:2010vc}. As we saw in~\cref{sec:power_spectra}, for a stationary and isotropic SGWB signal and stationary noise (that is uncorrelated with the signal), the auto- and cross-spectra of the different channels read
\begin{equation} 
 \langle \tilde d_i \tilde{d}^*_j \rangle' = \mathcal{R}_{ij}  P_h (f)   +   N_{ij} (f) \;,
\label{eq:data_correlation_app_XYZ}
\end{equation} 
where the prime indicates that we have stripped the frequency $\delta$-function and the factor $1/2$, and we have used \cref{eq:singlesidedsps} and \cref{eq:singlesidednps}. Under the assumptions made in~\cref{sec:noise_and_signal_model} the three interferometers have identical properties, and the noise spectra and response functions obey
\begin{eqnarray}
N_{\text{XX}} &=& N_{\text{YY}} =N_{\text{ZZ}} \;, \qquad N_{\text{XY}} = N_{\text{YZ}} =N_{\text{XZ}}\;, 
 \label{eq:noiseassumpt1} \\
\mathcal{R}_{ \text{XX} } &=& \mathcal{R}_{\text{YY} }= \mathcal{R}_{ \text{ZZ} }\;, \qquad  \mathcal{R}_{ \text{XY} } = \mathcal{R}_{ \text{YZ} }= \mathcal{R}_{ \text{XZ} } \;.
\label{eq:noiseassumpt2} 
\end{eqnarray}
The resulting symmetry properties of $\langle \tilde d_i \tilde{d}^*_j \rangle'$ imply that it can be diagonalized by 
\begin{eqnarray} 
\tilde{d}_{\text{A}} &=& \frac{1}{\sqrt{2}} \left(\tilde{d}_{\text{Z}} - \tilde{d}_{\text{X}} \right) \;, 
\nonumber\\ 
\tilde{d}_{\text{E}} &=& \frac{1}{\sqrt{6}} \left( \tilde{d}_{\text{X}} - 2\tilde{d}_{\text{Y}} +\tilde{d}_{\text{Z}} \right)  \;, 
\nonumber\\ 
\tilde{d}_{\text{T}} &=&  \frac{1}{\sqrt{3}} \left( \tilde{d}_{\text{X}} + \tilde{d}_{\text{Y}} + \tilde{d}_{\text{Z}} \right)  \;,
\label{signal-AET}
\end{eqnarray} 
or any other transformation related by a cyclic permutation in $\text{X}$, $\text{Y}$, $\text{Z}$, where $\tilde{d}_{\text{A}}, \tilde{d}_{\text{E}}, \tilde{d}_{\text{T}}$ and $\tilde{d}_{\text{X}}, \tilde{d}_{\text{Y}}, \tilde{d}_{\text{Z}}$ are the data in the AET and XYZ streams, respectively, and we have suppressed the argument to streamline our notation. 

\begin{figure}
    \centering
    \includegraphics[width = 0.48 \textwidth]{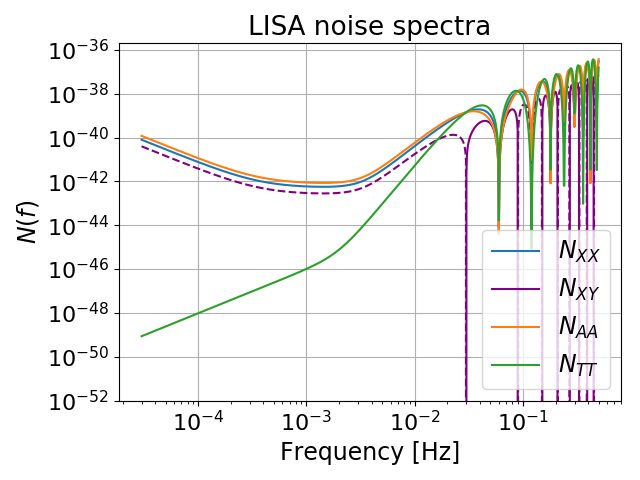}
    \includegraphics[width = 0.48 \textwidth]{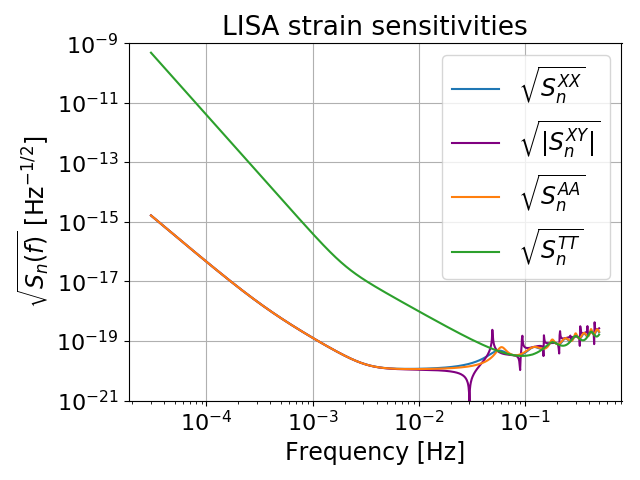}
    \caption{Left panel: LISA noise spectra in the XYZ and AET basis. Right panel: square root of the LISA strain noise for the different XYZ/AET combinations. \label{fig:PSD_strain} }
\end{figure}

The spectra in the AET basis become
\begin{eqnarray}
\label{eq:AA_EE}
\langle \tilde{d}_{\text{A}} \tilde{d}_{\text{A}}^* \rangle' = \langle \tilde{d}_{\text{E}} \tilde{d}_{\text{E}}^* \rangle'& =&  \left( \mathcal{R}_{ \text{XX} } - \mathcal{R}_{ \text{XY} }\right)  P_h(f) + N_{ \text{XX}} -  N_{ \text{XY}}   \;, \\ 
\label{eq:TT}
\langle \tilde{d}_{\text{T}} \tilde{d}_{\text{T}}^* \rangle' &=& \left( \mathcal{R}_{ \text{XX} } + 2 \mathcal{R}_{ \text{XY} } \right)  P_h(f) + N_{ \text{XX}} + 2 N_{\text{XY} }   \;,
\\
\langle \tilde{d}_{\text{A}} \tilde{d}_{\text{E}}^* \rangle' = \langle \tilde{d}_{\text{A}} \tilde{d}_{\text{T}}^* \rangle' = \langle \tilde{d}_{\text{E}} \tilde{d}_{\text{T}}^* \rangle'  &= & 0 \;.
\label{eq:AET_off_diag}
\end{eqnarray}
We see that the noise spectra and response functions in the AET basis are given by
\begin{gather}
\label{eq:responses_PSD}
N_{\text{AA}} = N_{\text{EE}}  = N_{\text{XX}} - N_{\text{XY}}\; , \qquad N_{\text{TT}} = N_{\text{XX}} + 2 N_{\text{XY}} \; ,\\ 
       \mathcal{R}_{ \text{AA}} = \mathcal{R}_{ \text{EE} } =  \mathcal{R}_{ \text{XX} }- \mathcal{R}_{ \text{XY}} \; , \qquad \mathcal{R}_{ \text{TT} } = \mathcal{R}_{ \text{XX} } + 2 \mathcal{R}_{ \text{XY} }\; .
\end{gather}
Substituting the noise auto- and cross-spectra,~\cref{eq:N_XX} and~\cref{eq:N_XY} into~\cref{eq:responses_PSD} we obtain
\begin{equation}
\begin{aligned}
    N_{ \text{AA} }(f, A, P)  & = N_{ \text{EE} }(f, A, P) =   \\
& =  8 \sin^2\left(\frac{2 \pi f L}{c}\right) \left\{ 4 \left[1 +\cos \left(\frac{2 \pi f L}{c} \right) + \cos^2 \left(\frac{2 \pi f L}{c} \right)\right] P_{\rm acc}(f, A) \;  + \right. \\ 
    & \hspace{3cm} \left. + \left[ 2 +\cos \left(\frac{2 \pi f L}{c} \right) \right]P_{\rm IMS}(f, P)  \right\}  \; , \\
\end{aligned}
\end{equation}
and
\begin{equation}
\begin{aligned}
    N_{ \text{TT} }(f, A, P) & =  16 \sin^2\left(\frac{2 \pi f L}{c}\right) \left\{ 2 \left[ 1 - \cos \left(\frac{2 \pi f L}{c} \right) \right]^2  P_{\rm acc}(f, A) \;  + \right. \\ 
    & \hspace{3cm} \left. + \left[ 1 - \cos \left(\frac{2 \pi f L}{c} \right) \right] P_{\rm IMS}(f, P) \right\} \; .
\end{aligned}
\label{eq:psdTT}
\end{equation}
These power spectrum densities are displayed in the left panel of~\cref{fig:PSD_strain}. 

Concerning the measurement of the noise amplitude parameters, from the left panel of~\cref{fig:PSD_comparisons} one might naively expect that data at $f \lesssim 2\times 10^{-3}$ contain precise information on the $A$ parameter, and in particular, to avoid signal contamination, $A$ would be measured most efficiently via the $\text{T}$ channel. However, the low-frequency expansion of \cref{eq:psdTT} yields 
\begin{equation}
  \hspace{-0.5cm}   N_{ \text{TT} }(f, A, P)  \simeq 8 \left( \frac{ f }{f_*}  \right)^2  \sin^2\left( \frac{ f }{f_*}  \right) \left[    \left(\frac{ f }{f_*} \right)^2  P_{\rm acc}(f, A) \; + P_{\rm IMS}(f, P) \right] \;,
\end{equation}
with $f_* \equiv (2 \pi L/c)^{-1} \simeq 0.019$\,Hz. So, at low frequencies, the contribution coming from $P_{\rm acc}$ is suppressed by a factor $(f/f_*)^2$ with respect to the contribution coming from $P_{\rm IMS}$, which means that the shape of $N_{\text{TT} }$ is dominated by $P_{\rm IMS}$ \emph{even at low frequencies}, as can be seen in the right panel of~\cref{fig:PSD_comparisons}. Despite this, the TT channel provides substantial information on $P_{\rm acc}$, constraining its amplitude well beyond the priors used in this paper (see~\cref{sec:data_analysis}).

\begin{figure}[htb]
    \centering
    \includegraphics[width = 0.48 \textwidth]{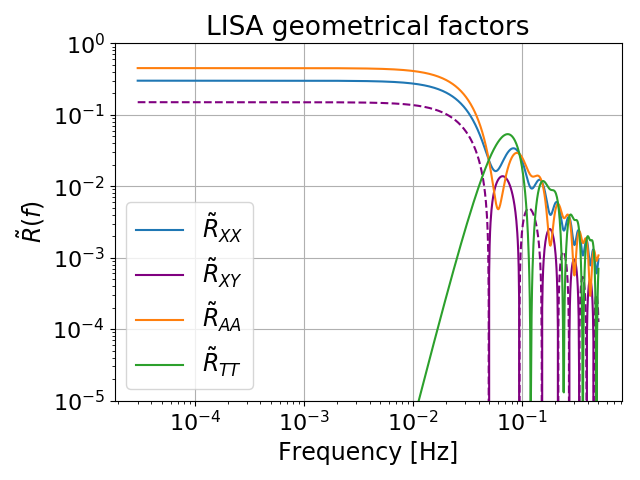}
    \includegraphics[width = 0.48 \textwidth]{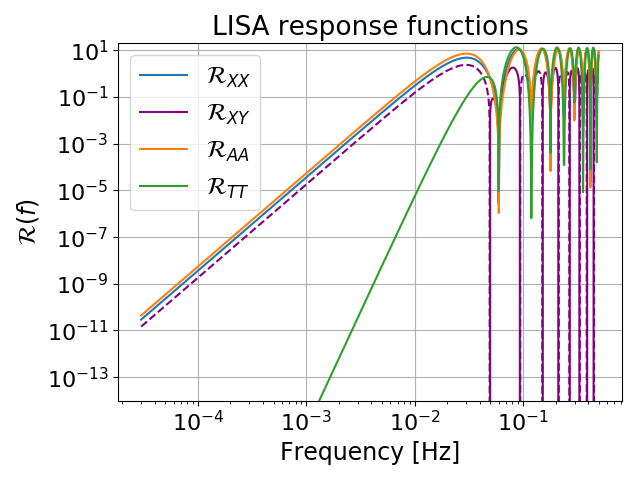}
    \caption{Left panel: geometrical contribution to the LISA response function. Right panel: full LISA response function.
    \label{fig:geometrical_response} }
\end{figure}

As we discuss in more detail in~\cref{sec:response}, the quadratic response functions for our choice of TDI variables can be written as
\begin{equation}
\mathcal{R}_{ij}(f) = 16 \sin^2\left(\frac{2 \pi f L}{c}\right) \left(\frac{2 \pi f L}{c} \right)^{2} \tilde{R}_{ij}(f) \; . \qquad 
\end{equation}
The first factor reflects the choice of TDI variables, the factor $2\pi f L/c$ appears because LISA measures frequency changes rather than time delays (see~\cref{sec:formulae} for details), and $\tilde{R}_{ij}$ is a factor that depends on the geometry of the detector. These geometrical factors and the corresponding response functions are shown in fig.~\ref{fig:geometrical_response}.

In practice, we work with the numerical expressions shown in~\cref{fig:geometrical_response}, but for some purposes simple analytic approximations are helpful. For the XX channel (as well as the YY and ZZ channels), the geometrical factor is well approximated by~\cite{Cornish:2018dyw}
\begin{equation}
	\tilde{R}_{ \text{XX} }(f) = \frac{3}{10}\frac{1}{1 +0.6 \left(\frac{2 \pi f L}{c}\right)^2} \; . 
\end{equation}
Similarly, for the $\text{AA}$ channel (as well as the $\text{EE}$ channel) and for the $\text{TT}$ channel, we find
\begin{eqnarray}
	\tilde{R}_{ \text{AA} }(f) &=& \frac{9}{20}\frac{1}{1 +0.7 \left(\frac{2 \pi f L}{c}\right)^2} \,,\\ \tilde{R}_{ \text{TT} }(f) &=& \frac{9}{20}\frac{\left(\frac{2 \pi f L}{c}\right)^6}{1.8 \times 10^3 +0.7 \left(\frac{2 \pi f L}{c}\right)^8} \,. 
\end{eqnarray}

From the power spectrum density and the response functions we can define the strain sensitivities
\begin{equation}
		S_{n\,ij}(f, A, P ) = \frac{N_{ij}(f,A,P)   }{  \mathcal{R}_{ij}(f)  }   = \frac{N_{ij}(f,A,P)   }{   16 \sin^2\left(\frac{2 \pi f L}{c}\right) \left(\frac{2 \pi f L}{c} \right)^{2} \tilde{R}_{ij}(f)} \; .
\end{equation}
Plots of these quantities are shown in the right panel of~\cref{fig:PSD_strain}.

Putting everything together, we arrive at the noise model used for the generation of the noise realizations and extraction of noise parameters used in our analyses, in units of the energy density parameter
\begin{equation}
    \label{eq:noise_model}
	\Omega_{n, ij} h^2 (f, A, P) \equiv \frac{4 \pi^2 f^3}{3 (H_0/h)^2} S_n^{ij}(f, A, P) \; ,
\end{equation}
where $H_0/h \simeq 3.24 \times 10^{-18}$ 1/s. 

\begin{figure}
    \centering
    \includegraphics[width = 0.48 \textwidth]{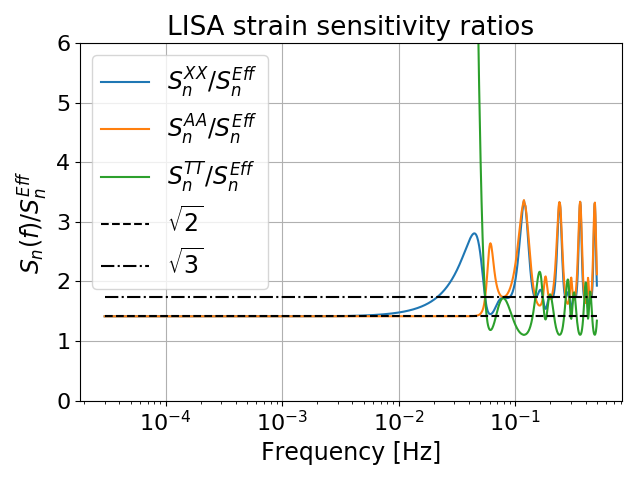}
    \includegraphics[width = 0.48 \textwidth]{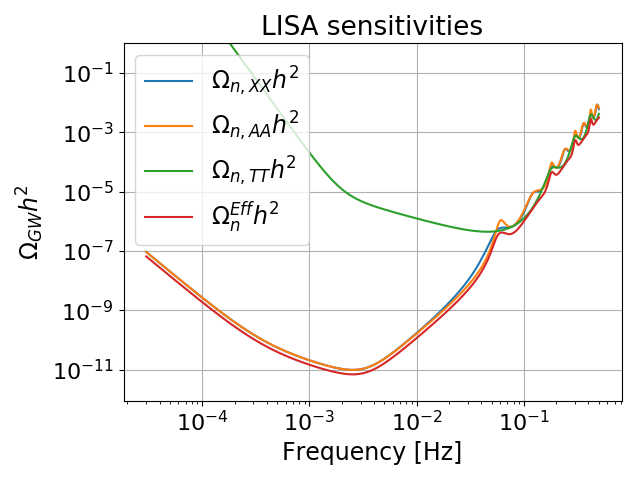}
    \caption{Left panel: ratios between the strain sensitivities for the different channels and the effective strain defined in~\cref{eq:effective_strain}. Right panel: LISA sensitivity (see~\cref{eq:noise_model}) for the different channels and for their inverse variance weighted combination. \label{fig:PSD_ratios} }
\end{figure}

\subsection{Optimal channel combination}
\label{sec:optchannel}
Since both the signal and noise covariance in the AET basis are diagonal, it is also straightforward to form an optimal (unbiased and with minimal variance) power spectrum estimator from the three channels~\cite{aet}. Reserving the lower case latin indices for formulae that apply for both the XYZ and the AET basis, and introducing greek indices for the AET basis, $\alpha\in\{\text{A},\text{E},\text{T}\}$, we can define the time stream variables 
$
\hat{d}_\alpha = \tilde{d}_\alpha/\sqrt{\mathcal{R}_{\alpha\alpha}}\,,
$
with unit response to the SGWB signal
\begin{equation}
\langle\hat{d}_{\alpha} \hat{d}_{\alpha}^* \rangle' =    P_h(f) + S_{n\,\alpha} \;,
\end{equation}
where $S_{n\,\alpha}=N_{\alpha\alpha} / R_{\alpha\alpha}$
is the effective noise power spectral density or sensitivity. In terms of these variables the unbiased power spectrum estimator for a single channel is 
$
    \hat{P}_\alpha=|\hat{d}_\alpha|^2-S_{n\,\alpha}\,,
$
and we can define an unbiased estimator by combining the channels 
\begin{equation}
    \hat{P}=\sum_\alpha w_\alpha \hat{P}_\alpha\,,\qquad\text{with}\qquad \sum_\alpha w_\alpha = 1\,.
\end{equation}
As usual, the optimal linear combination is obtained by inverse variance weighting
\begin{equation}
    w_\alpha = \frac{(P_h+S_{n\,\alpha})^{-2}}{\sum\limits_\alpha (P_h+S_{n\,\alpha})^{-2}}\,.
\end{equation}
By calculating the variance of this estimator, we see that in the limit of small signal-to-noise ratio per mode, the effective noise power spectral density for this combination of channels is given by
\begin{equation}
    \label{eq:effective_strain}
    S_n = \frac{1}{\sqrt{\sum\limits_\alpha S_{n\,\alpha}^{-2}}}\,.
\end{equation}
A comparison between this quantity and the standard noise spectra is presented in~\cref{fig:PSD_ratios}. The effective sensitivity for the optimal combination provides a simple way to compare sensitivities for a single channel as used in ref.~\cite{Caprini:2019pxz} and three channels. This is shown in~\cref{fig:PSD_ratios}. Beyond this comparison, we will not work with this combination and will continue to use the XYZ and AET bases, which, as we will see, has practical advantages. However, this combination would be a natural choice to visually present the results of a measurement of a SGWB with LISA.

\section{Data analysis techniques and methodology}
 \label{sec:data_analysis}
 In this section we discuss the data analysis techniques employed in this work. As usual, we compute the posterior probability for the model parameters using the Bayes theorem:
\begin{equation}
	p(\vec{\theta},\vec{n}|D) = \frac{\pi_S(\vec{\theta})  \pi_N(\vec{n}) \mathcal{L}(D | \vec{\theta},\vec{n})} {\mathrm{Z}_{N,S} (D) }\; ,
\end{equation}
where  $ \mathcal{L}(D|\vec{\theta},\vec{n}) $ is the likelihood of the data $D$, $\pi_N$ and $\pi_S$ are the prior for the noise parameters $\vec{n}$ and for the signal parameters $\vec{\theta}$, respectively, and $\mathrm{Z}_{N,S} (D) $ is the model evidence. Since we are not interested in the model evidence (or the normalization of the posterior probability in general), we work with the unnormalized posterior $P(\vec{\theta},\vec{n}|D)$. Omitting $D$ for simplicity, we have
\begin{equation}
  P(\vec{\theta},\vec{n}) = 
  \pi_S(\vec{\theta}) \pi_N(\vec{n}) \mathcal{L}(D | \vec{\theta},\vec{n})
  \propto p(\vec{\theta},\vec{n}|D) \; .
\end{equation}
In this work (consistent with the analysis in ref.~\cite{Caprini:2019pxz}) we assume a Gaussian prior for the noise parameters:
\begin{equation}
  \vec{n} = (A, P) \sim \mathcal{N}\left(\vec{\mu}, \mathbf{\Sigma}\right)
  \,,\quad\text{with}\quad
  \vec{\mu} = (\bar{A}, \bar{P}) = (3, 15)
  \,,\quad\text{and}\quad \mathbf{\Sigma} = \mathrm{diag}(0.2\,\vec{\mu})
  \; .
\end{equation}
The prior on the signal parameter is model-dependent, but in all subsequent analyses in this paper will be taken as uniform (or log-uniform, for parameters representing amplitudes).

The Maximum A Posteriori (MAP) parameters (denoted with $\vec{\theta}_b$ and $\vec{n}_b$) are defined as
\begin{equation}
    \label{eq:MLEestimation}
 	\left. \partial_j \ln P(\vec{\theta},\vec{n}) \right|_{\vec{\theta}_b , \vec{n}_b } \equiv 0 \; , 
\end{equation}
where the $j$ index defines one equation per parameter, both signal and noise ones. As it is customary, the Fisher information matrix is defined as
\begin{equation}
\mathcal{I}_{ij} = C_{ij}^{-1}\equiv - \langle \left. \partial_i \partial_j \ln P(\vec{\theta},\vec{n})   \right|_{\vec{\theta}_b, \vec{n}_b} \rangle \; .
\end{equation}
Notice that with this definition the Fisher matrix also contains information about the priors.

\subsection{Data generation and likelihood}
\label{sec:datagenlike}
For our simulations, we will assume that the data is taken during a full mission duration of $4$ years with $75\%$ efficiency, giving an effective observation time of $3$ years. As discussed in~\cref{sec:power_spectra}, we will assume the TDI data to be divided into roughly $95$ \textit{chunks} (or data segments) of $11.5$ days each, clean of noise glitches and transient signals. Since the frequency range for LISA approximately extends from $3 \times 10^{-5}$ Hz to $5 \times 10^{-1}$ Hz, this amounts to around $5 \times 10^{5}$ data points per frequency, with a frequency resolution of $\approx 10^{-6}$ Hz, for a total of roughly $5 \times 10^{7}$ data points.

For each chunk $l$ and channels $i,j$, we generate the data $D_{ij}^{l}(f)=d_{i}^l(f) d^{l \, *}_{j}(f)$ directly in the frequency domain, following the procedure described in ref.~\cite{Caprini:2019pxz}. Generation in the frequency domain allows us to ignore window effects and overlapping segments, simplifying the process.

We compress the simulated data by averaging over the chunks and binning the data. Specifically, we define the average
    \begin{equation}
        \bar{D}_{ij}(f) \equiv \frac{1}{N_{\text{c}}} \sum_{l=1}^{N_\text{c}} D_{ij}^{l}(f) \; , 
    \end{equation}
where $N_{\text{c}} = 95$ is the number of data segments. We then bin the data in the frequency domain into a new data set $\left(f_{ij}^{(k)}, \mathcal{D}_{ij}^{(k)}\right)$. In practice, we use inverse variance weighting 
    \begin{eqnarray}
        f_{ij}^{(k)} & \equiv &
        \sum_{f \in \left\{f^{(k)}\right\}} w_{ij}^{(k)}(f)\,f   \; ,  \\
        \mathcal{D}_{ij}^{(k)} & \equiv &
       \sum_{f \in \left\{f^{(k)}\right\}}  w_{ij}^{(k)}(f)\,\bar{D}_{ij}(f)\; ,
    \end{eqnarray}
    where $\left\{f^{(k)}\right\}$ are the (discrete) frequencies in bin $k$, and the weights are 
    \begin{equation}
    w_{ij}^{(k)}(f) =  \frac{\sigma^{-2}_{ij}(f)} 
        { \sum\limits_{f \in \left\{f^{(k)}\right\}} \sigma^{-2}_{ij }(f) } \; , 
    \end{equation}
where $\sigma^2_{ij}(f)$ is an estimate of the variance of the segment-averaged data $\bar{D}_{ij}(f)$. For the analyses presented in this work we assume the variance to be well approximated by LISA's sensitivity introduced in~\cref{eq:noise_model}.\footnote{It can be shown that this procedure would be optimal in the case of noise-dominated data with real noise parameters equal to their face values. In the case of a SGWB detection with signal-to-noise ratio comparable to or larger than unity per mode, we would have to update the variance of the data to account for it.} Since the variance depends on the combination of channels, the inverse variance weighting implies that the discrete frequencies also depend on the channel combination.
    
In practice, the second step is performed by requiring a maximum of 1000 linearly spaced points per frequency decade. For different bins the coarse-grained points are obtained by summing over different numbers of data points. To keep track of the weights carried by the different coarse-grained data points, we denote as $n_{ij}^{(k)}$ the number of points within bin $k$ for the cross-spectrum of channels $i$ and $j$.

As a starting point to build a likelihood for the data, it is tempting to assume a simple Gaussian likelihood
\begin{equation}
    \label{eq:gaussian_likelihood}
	\ln \mathcal{L}_G(D|\vec{\theta},\vec{n}) = - \frac{ N_{\text{c}} }{2} \sum_{i,j} \sum_{k} n_{ij}^{(k)} \left[ \frac{ \mathcal{D}_{ij}^{th}(f_{ij}^{(k)}, \vec{\theta}, \vec{n}) -\mathcal{D}_{ij}^{(k)}}{ \mathcal{D}_{ij}^{th}(f_{ij}^{(k)}, \vec{\theta}, \vec{n}) } \right]^2 \; , 
\end{equation}
where the indices $i,j$ run over the different channel combinations, the index $k$ runs over the coarse-grained data points, and we have defined $ \mathcal{D}_{ij}^{th}(f, \vec{\theta}, \vec{n}) \equiv  \mathcal{R}_{ij}  \Omega_{GW}h^2(f, \vec{\theta}) +  \Omega_{n, ij} h^2 (f, \vec{n})  $ as our model for the data. Notice that since in the XYZ basis the channels are correlated, $\mathcal{D}_{ij}$ is not diagonal in channel space. On the other hand in the AET basis this likelihood reduces to the sum of the diagonal elements in channel space.

However, it is known that the Gaussian likelihood in~\cref{eq:gaussian_likelihood} would give a result which is systematically biased~\cite{Bond:1998qg, Sievers:2002tq, Verde:2003ey, Hamimeche:2008ai}, due to the mild non-Gaussianity of the full likelihood of the non-averaged data. In order to correct for this bias we also introduce a log-normal likelihood~\cite{Bond:1998qg, Sievers:2002tq} 
\begin{equation}
\label{eq:lognormal_likelihood}
\ln \mathcal{L}_{LN}(D|\vec{\theta},\vec{n}) = - \frac{N_{\text{c}}}{2} \sum_{i,j} \sum_{k} n_{ij}^{(k)}  
\ln ^2 \left[ \frac{ \mathcal{D}_{ij}^{th}(f_{ij}^{(k)}, \vec{\theta}, \vec{n})  }{ \mathcal{D}_{ij}^{(k)} } \right]  \; , 
\end{equation}
and define our final likelihood as~\cite{Verde:2003ey}
\begin{equation}
\ln  \mathcal{L} = \frac{1}{3} \ln \mathcal{L}_{G}+ \frac{2}{3} \ln  \mathcal{L}_{LN} \; .
\end{equation}
It can be shown~\cite{Verde:2003ey} that this likelihood accounts for the skewness in the full non-Gaussian likelihood, giving a more accurate result for the model parameters.

As we just saw, the likelihood depends on a theoretical spectrum. This could be a model for one or several sources of a SGWB. Here we instead employ a non-parametric approach which aims to reconstruct the frequency dependence of the SGWB without assuming any previous knowledge of its spectral shape. We define a piece-wise model for the signal on a set of frequency intervals (bins). Within each of these bins the signal is assumed to be well approximated by a power law:
\begin{equation}
    \label{eq:signal_model}
h^2 \Omega_{\rm GW} \left( f ,\, \vec{\theta}_i \right) = \sum_i \;  10^{\alpha_i}  \, \left( \frac{f}{\sqrt{f_{{\rm min},i} \, f_{{\rm max},i}}} \right)^{n_{t, i} }  
 \; \Theta \left( f - f_{{\rm min},i} \right) \,  \Theta \left( f_{{\rm max},i} - f \right) \,, 
\end{equation} 
 where $\Theta$ is the Heaviside step function, the index $i$ runs over the bins, $f_{\rm min, i}$ and $f_{\rm max, i }$ are respectively the minimum and maximum frequencies in bin $i$, $f_{*,i} \equiv \sqrt{f_{{\rm min},i} \, f_{{\rm max},i}}$ is the pivot frequency for bin $i$ and $\vec{\theta}_i = \{ \alpha_i, n_{t,i} \}$ are the signal parameters for bin $i$ (respectively denoting amplitude at the pivot, and tilt). As we discuss below, the number of bins will be adjusted dynamically.

\subsection{Algorithm for binned reconstruction of a SGWB signal}
\label{sec:sgwbinner_algorithm}
Here we present an update of the algorithm described in ref.~\cite{Caprini:2019pxz}, and implemented in the \texttt{SGWBinner} code, to reconstruct the frequency shape of the SGWB, using the binned power-law signal model and the noise model presented in \cref{sec:noise_and_signal_model}. Given some pre-determined binning, this problem reduces to finding the solution of eq.~\eqref{eq:MLEestimation} over all parameters simultaneously (the two signal parameters per bin, and the two noise parameters which are common to all the bins) together with some estimation of the uncertainty of the reconstruction.

This non-trivial minimization problem may be tackled with Monte Carlo (MC) methods. However, if we wish to leave the number of bins as a free parameter, a MC sample would have to be obtained for each binning configuration, and then the optimal number of bins would be chosen by e.g.~comparing Bayesian evidences of different bin numbers. This process is computationally expensive. For this reason, similarly to ref.~\cite{Caprini:2019pxz}, we adopt a different strategy:
\begin{enumerate}
    \item We build a prior for the noise parameters using the TT-channel (more on this below).
    \item We bin the signal in frequency in a given initial number of bins, which will be the maximum allowed for this run.
    \item We minimise the posterior for channel AA independently for each bin, imposing the TT-defined noise prior, for the two independent signal and two independent noise parameters.
    \item For all pairs of neighboring bins, we iteratively check whether merging the two bins is statistically favoured, using the Akaike Information Criterion~\cite{Aka74} as described in ref.~\cite{Caprini:2019pxz}. This procedure reduces the chance of overfitting, improving the quality of the reconstruction. If one or more merges are deemed favourable, we re-define the bins and go back to the minimisation step.
    \item When the number of bins has converged, we estimate the error on the reconstruction and provide a visualization of the $1\sigma$ and $2\sigma$ regions for the reconstructed signal, using the procedure described in~ref.~\cite{Caprini:2019pxz}.
\end{enumerate} 
The advantage of this methodology is that it greatly simplifies the problem from the numerical point of view, making it easily solvable in a reasonable amount of computation time. The first four steps are indeed fast. We can run them several times for different initial numbers of bins, and ultimately proceed with the final step only for the binning configuration that leads to the best Akaike Information Criterion.

At this point we should stress that step 1 is crucial for two main reasons: 
\begin{itemize}
    \item A wrong estimate of the noise parameters may bias the signal reconstruction.
    \item Since the noise parameters are actually common to all bins, this procedure may only be trusted if the results obtained in the per-bin analysis are consistent among all bins.
\end{itemize}
In ref.~\cite{Caprini:2019pxz}, in order to define suiting priors for the noise parameters, we took advantage of the morphology of the spectra of the data. Based on the reasonable assumption that at very low and very high frequencies the sensitivity of the detector is significantly weaker than at the central $\mathcal{O}(10^{-3})$ Hz region, we used these outer frequency regions to define strong priors on the instrument noise for the interval of the frequency range in which we attempt to reconstruct the signal.

That procedure provided good results when working with a single TDI channel. In the present work, however, we work with the noise-orthogonal A, E, and T TDI combinations~\cite{aet}. As explained in~\cref{sec:noise_and_signal_model}, at low frequencies the TT channel combination would greatly suppress any GW signal, and would therefore allow us to use it for a definition of the instrument noise prior density, without the need to limit the frequency range on which we reconstruct the background signal. In practice, we find the posterior in the TT channel combination of a power-law signal and the noise parameters, and build a prior on the $A$ and $P$ noise parameters by marginalizing over the rest. This is one of the main improvements of the present analysis with respect to the one of ref.~\cite{Caprini:2019pxz}. The posterior on the $A$ and $P$ noise parameters provided respectively by the left- and right-most bins are a good approximation to the posterior obtained from a joint fit to all bins. The noise parameters in the inner bins will be less constrained, but, as explained above, kept consistent thanks to the prior placed on them by the TT channel data.

As an optional final step in the algorithm, once the optimal number of bins has been set, we may then run an MC sampler on the total posterior for all bins and all channel combinations. We have implemented this step in the \texttt{SGWBinner} code by interfacing with the \texttt{Cobaya} MC framework \cite{Torrado:2020dgo}. One would expect this procedure to provide a more accurate estimation of the uncertainty on the reconstruction, but in practice we found little difference with the results of the original algorithm, as discussed in \cref{sec:mccomp}.

\section{SGWB signal reconstruction}
\label{sec:results}
In this section we apply the extension of the \texttt{SGWBinner} algorithm described in the previous section to a number of mock data sets generated as described in~\cref{sec:datagenlike} in which we have injected large-amplitude SGWB signals. The aim of this section is to asses the capability of our pipeline to perform a model-independent reconstruction of SGWB signals with distinct frequency dependence and make a comparison with the reconstruction using one LISA single-channel that was done in ref.~\cite{Caprini:2019pxz}. In all the examples in this section, we have started from log-spaced bins (as opposed to equal-SNR bins).

\subsection{Power-law reconstruction}
\label{sec:PL}
Our first benchmark signal is a simple power-law described by
\begin{equation}
   h^2 \Omega_{\rm GW} \left( f \right)  =10^{\alpha_{*}}\left( \frac{f}{0.001 \, {\rm Hz}} \right)^{n_t}
\end{equation}
where we have chosen an amplitude $\alpha= -11.352$ at the pivot frequency $f_* = 0.001 $\,{\rm Hz}  and a tilt $n_t = 2/3$. A power-law spectrum with a similar slope can be used to approximate some inflationary models~\cite{Bartolo:2016ami} as well as the astrophysical SGWB from stellar-origin black hole binaries and neutron star binaries~\cite{LIGOScientific:2019vic}  (see \cref{sec:foreground} for more details).

\begin{figure}[t]
    \centering \includegraphics[width=0.95\textwidth]{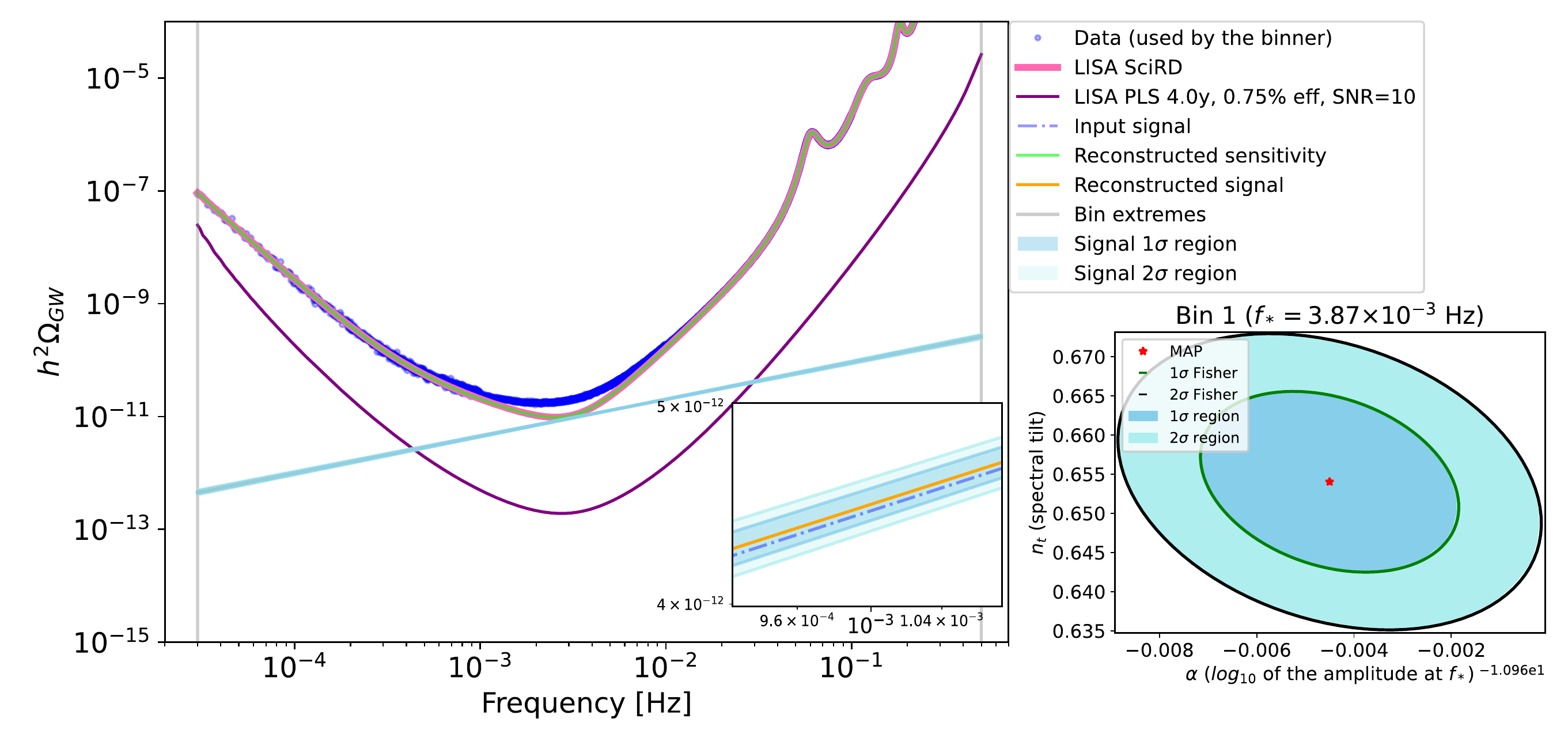}
  \caption{Reconstruction of the power-law signal described in~\cref{sec:PL} in the AET channels using 10 initial bins, which converged to a single one. On the bottom right, we show the contour plot of the posterior of the parameters of the single bin, including the Maximum A Posteriori (MAP). }
    \label{fig:plrec}
\end{figure}

The result of the reconstruction algorithm is shown in~\cref{fig:plrec}. Due to the smooth nature of the signal, we started with a small number of bins, 10, which, as expected being the signal a smooth power law across the whole LISA band, converged to a single bin and produced a successful reconstruction. We also show the $1\sigma$ and $2\sigma$ posterior contours of the amplitude and slope of the reconstructed signal in the single bin. When translated from the pivot scale of the power law to the pivot of the single bin, the amplitude and the slope fall comfortably within the 1-$\sigma$ C.L.\ contour. The SNR of the reconstructed signal is approximately $483$. The reconstructed signal (yellow line) matches the input signal (dashed-dot blue) very well within the error bars quantified by the thickness of the light-blue curves, as can be seen in the zoomed region around the pivot in~\cref{fig:plrec}.

\begin{figure}[ht]
    \centering
    \includegraphics[width=0.85\textwidth]{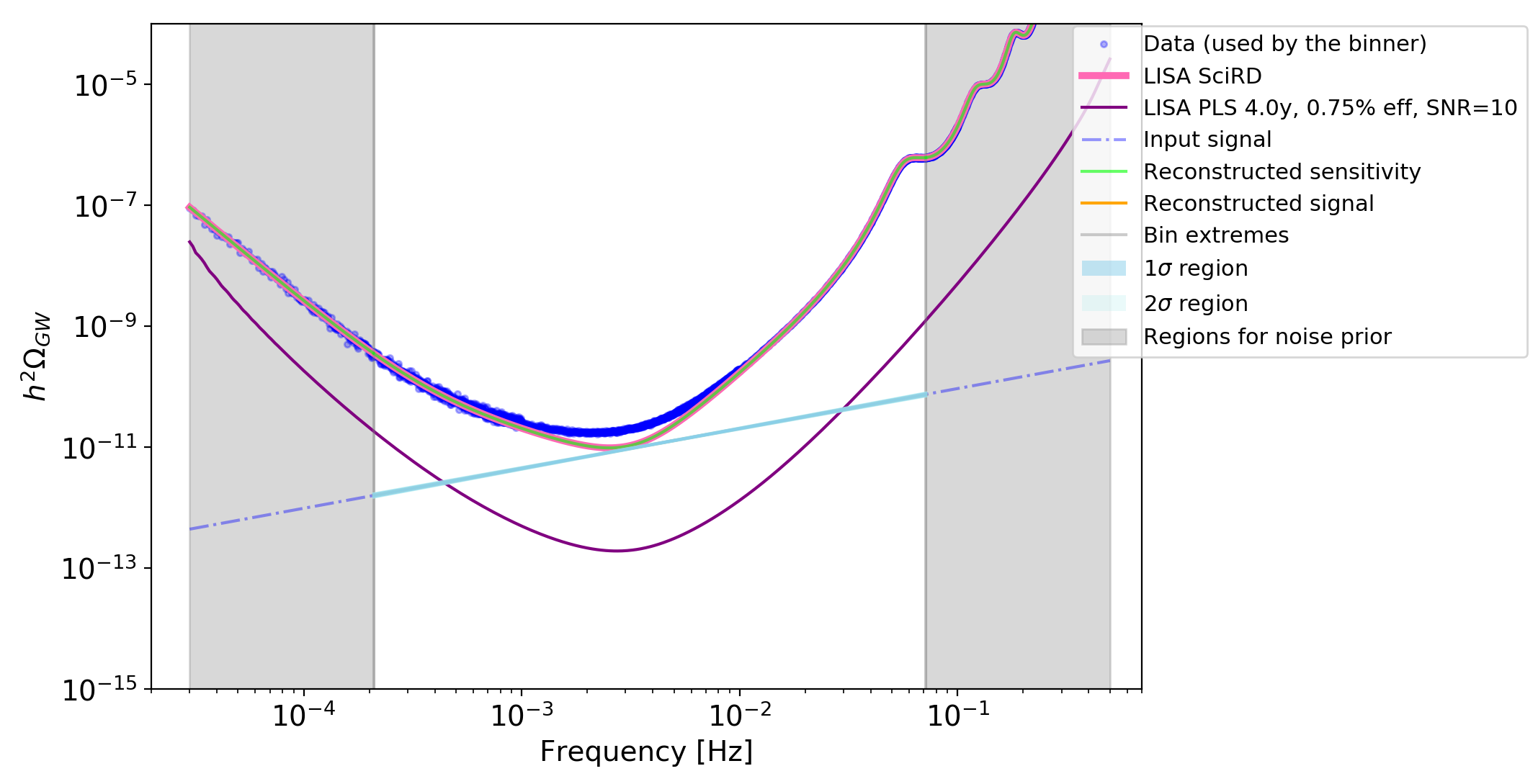}
    \caption{Reconstruction of the power-law signal in~\cref{fig:plrec} using X-channel data only. Notice that there is no reconstruction in the outer bins, which must be used to estimate the noise.}
    \label{fig:powerlawxyz}
\end{figure}

For the power-law case we test the procedure also in the X channel for a comparison. The result of the reconstruction with the same input parameters is given in~\cref{fig:powerlawxyz}. In this case we also recover the power-law parameters at 1-$\sigma$, and the SNR of the reconstructed signal is approximately $474$. The reconstructed signal here is also very well within the error bars of the reconstruction. The error bar of a reconstruction performed using the three TDI 1.5 channels would be $\sim\sqrt{2}$ narrower than that of the single-channel case, as discussed in~\cref{sec:optchannel}.

\subsection{Broken power-law reconstruction}
\label{sec:Broken_PL}
\begin{figure}[t!]
  \begin{subfigure}{0.95\textwidth}
    \centering \includegraphics[width = 0.8 \textwidth]{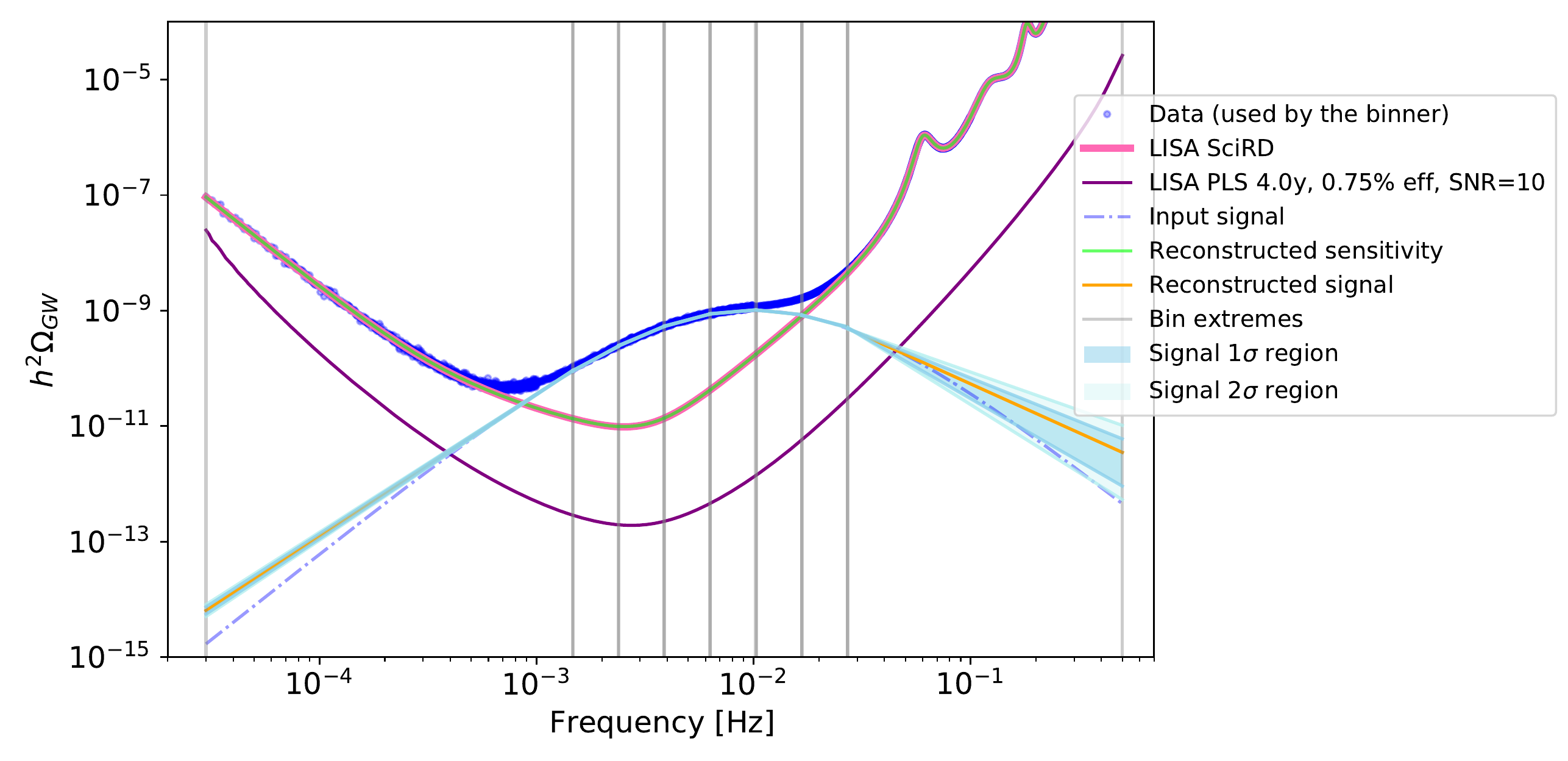}
    \caption{}
    \label{fig:bplrec}
  \end{subfigure}
  \begin{subfigure}{0.95\textwidth}
    \centering
    \begin{tabular}{c@{\hspace{0mm}}c@{\hspace{0mm}}c}
    \includegraphics[width=0.32 \textwidth]{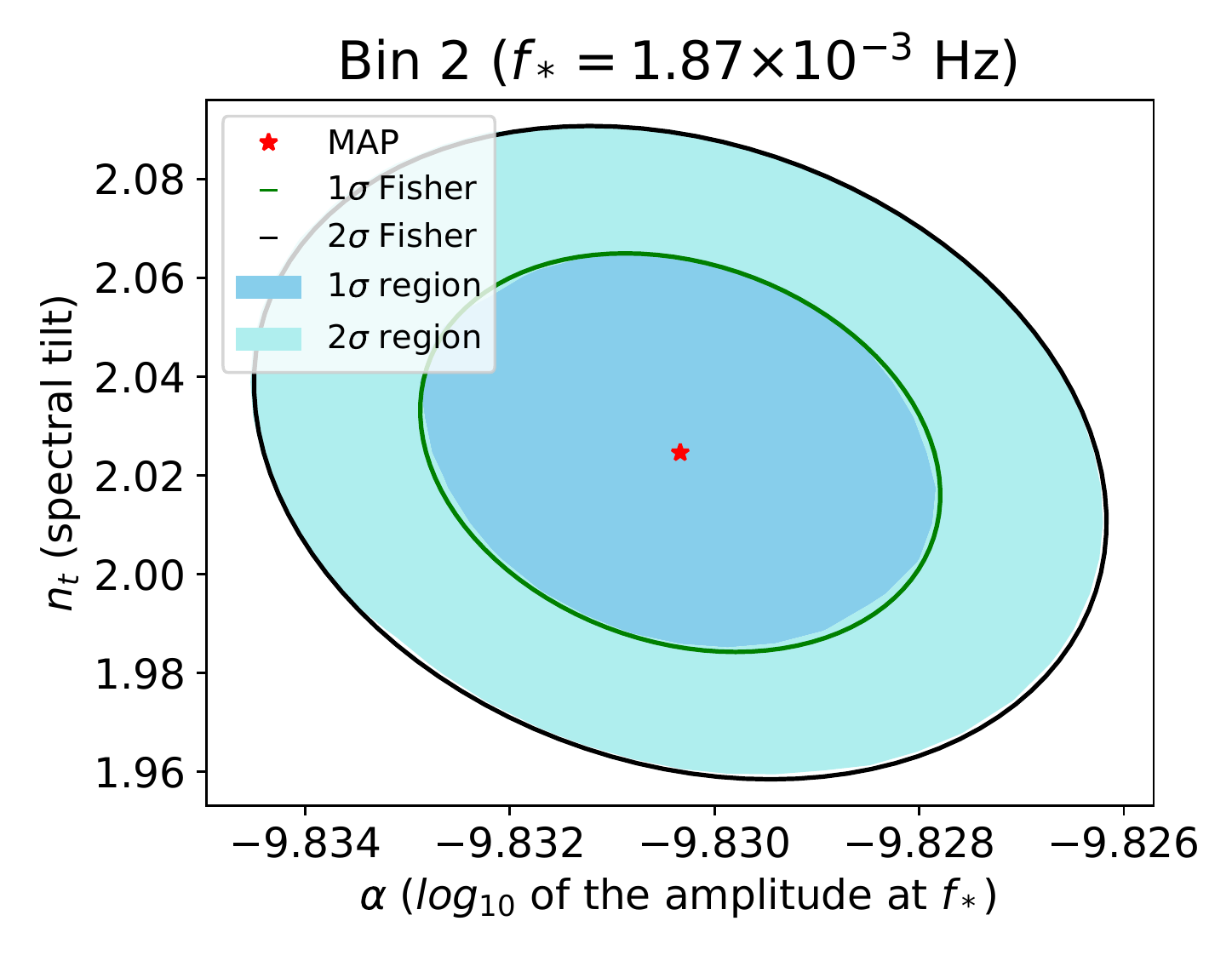}&
    \includegraphics[width=0.32 \textwidth]{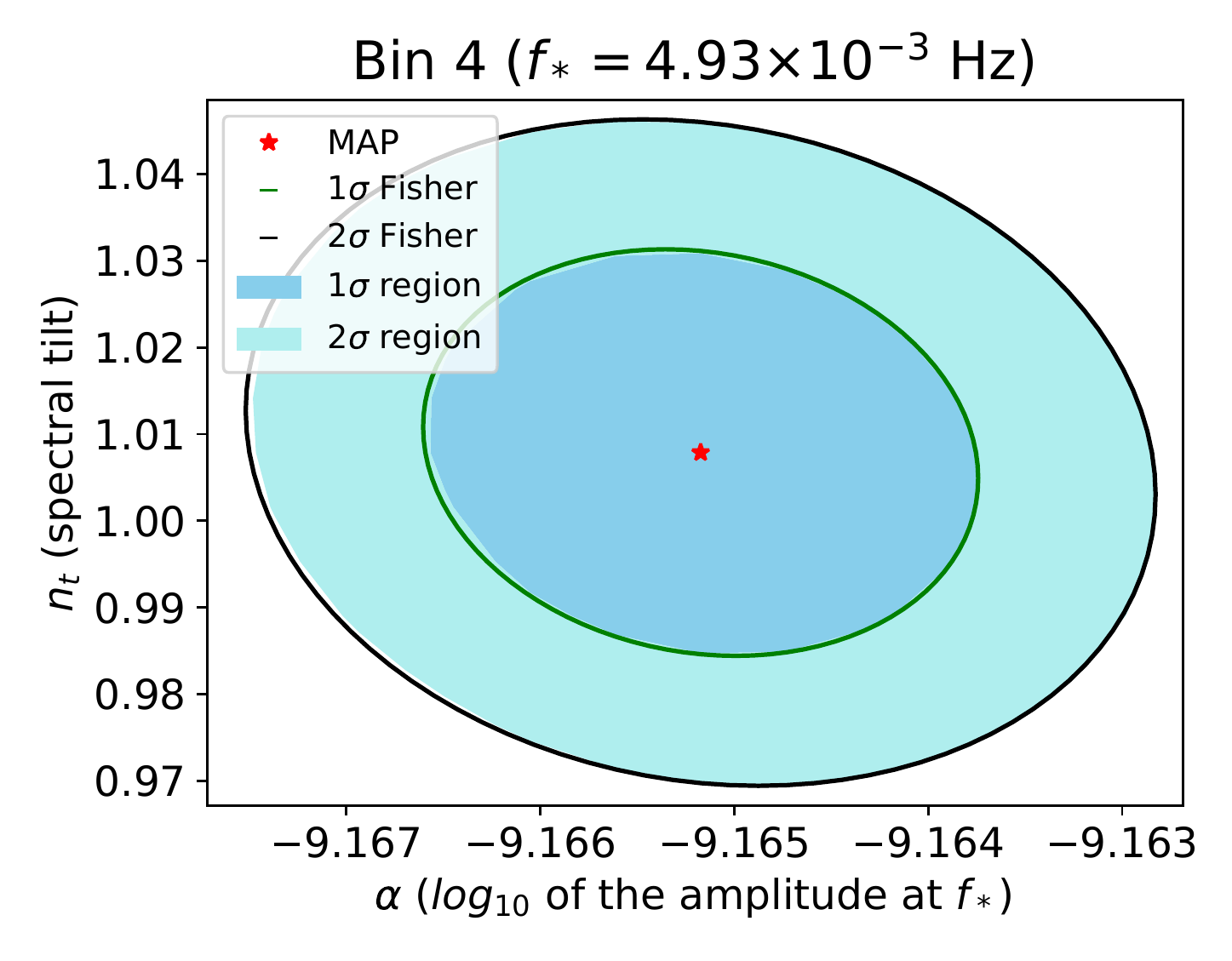}&
    \includegraphics[width=0.32 \textwidth]{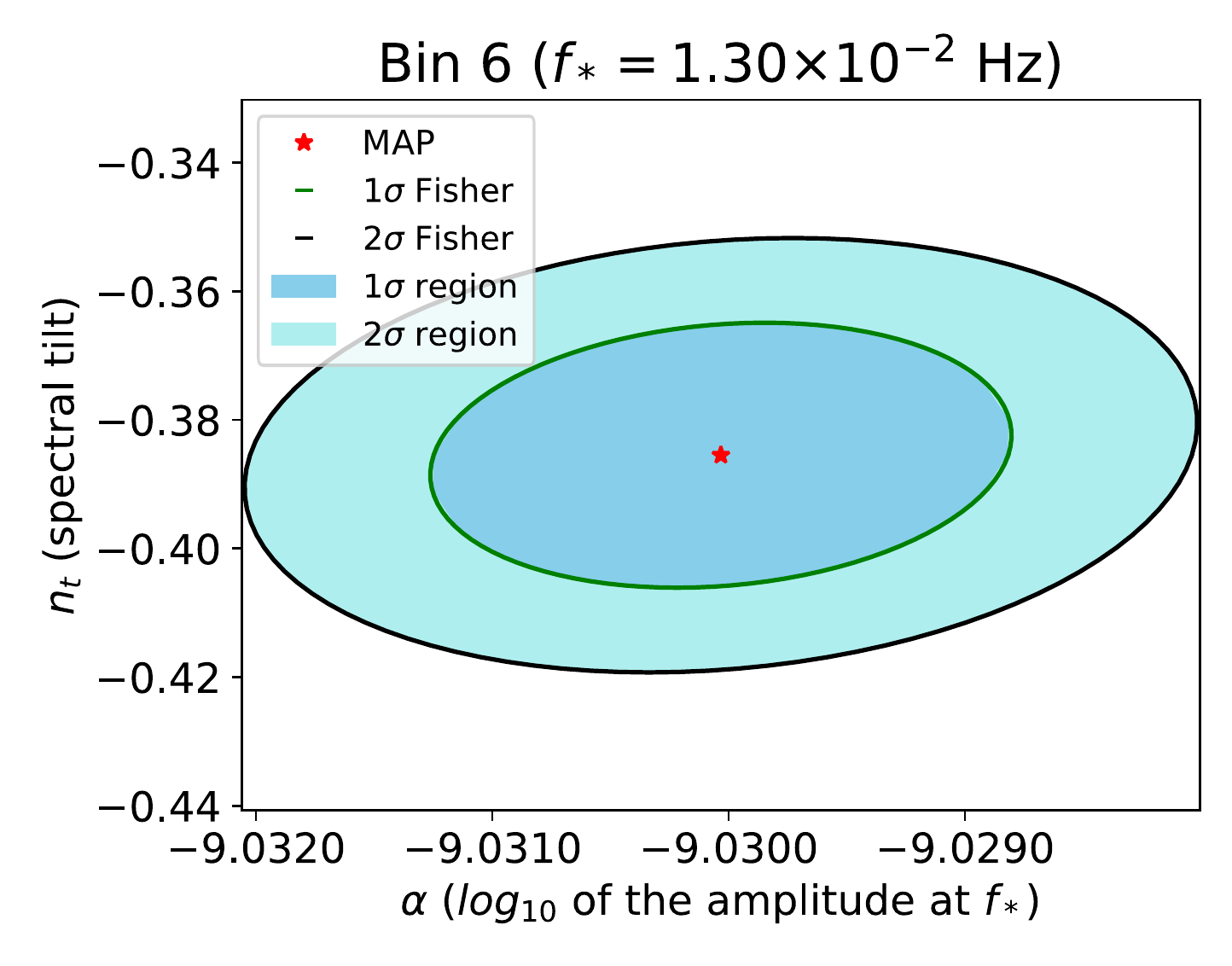}\\
    \includegraphics[width=0.32 \textwidth]{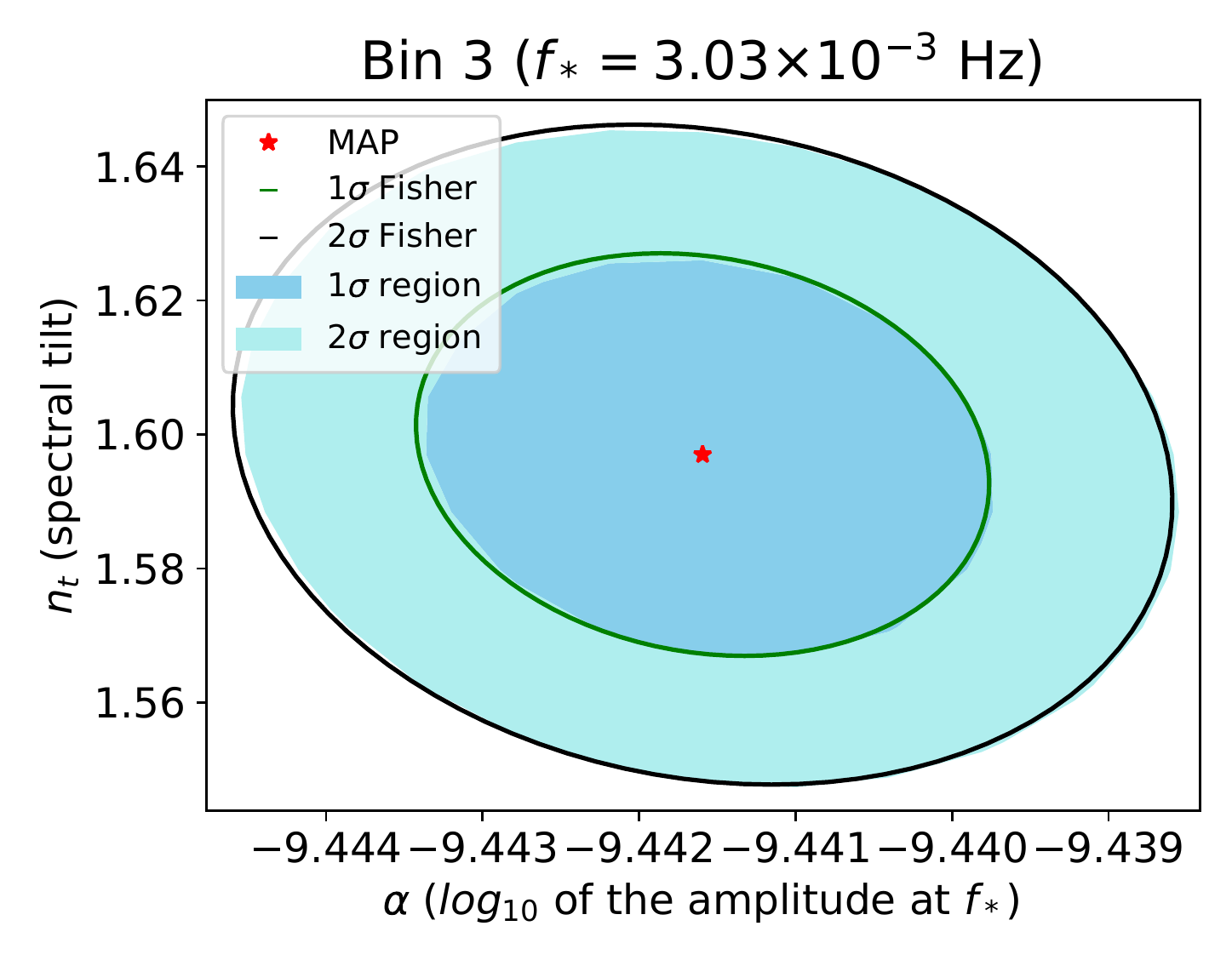}&
    \includegraphics[width=0.32 \textwidth]{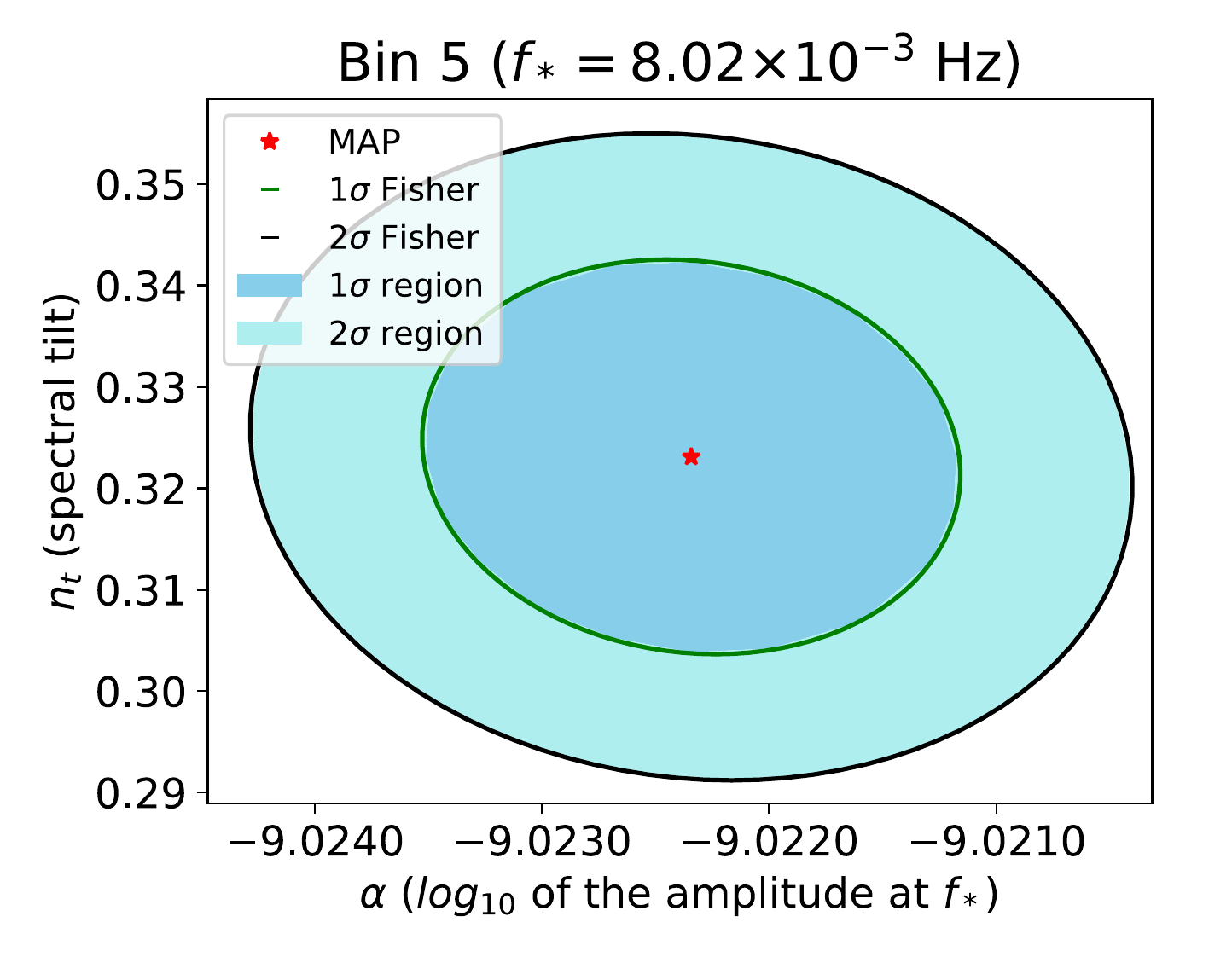}&
    \includegraphics[width=0.32 \textwidth]{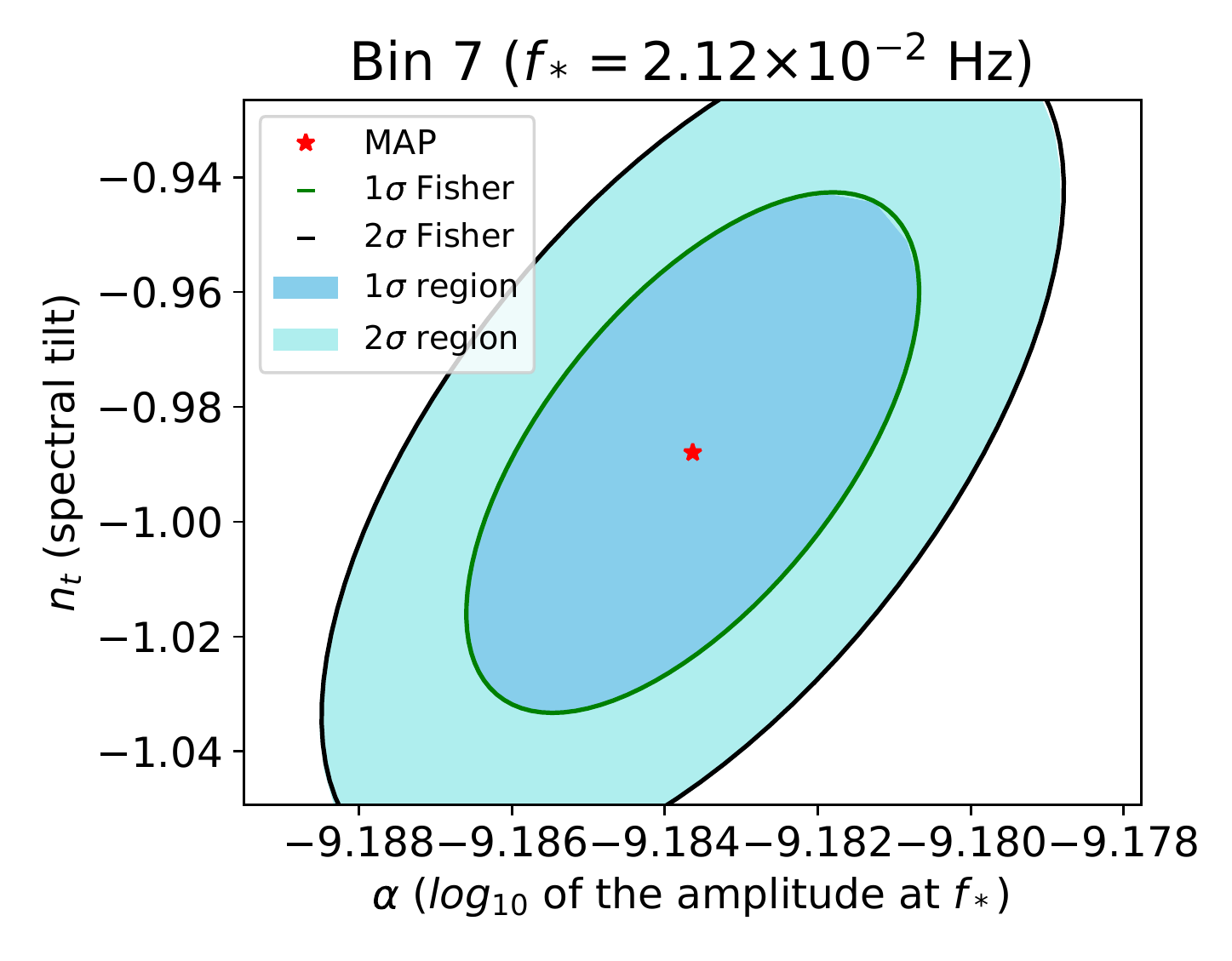}
    \end{tabular}
    \caption{}
    \label{fig:bplcp}
  \end{subfigure}
  \caption{\textbf{(a)} Reconstruction of the broken power-law signal described in~\cref{sec:Broken_PL} in the AET channels using 20 initial bins, which converged to 8. \textbf{(b)} Contour plots of the posterior of the parameters of the inner bins, showing the maximum a posteriori (MAP).}
  \label{fig:bpl}
\end{figure}

Our second benchmark signal is a broken power law with the following functional form:
\begin{equation}
    \label{eq:Broken_pl}
     h^2 \Omega_{\rm GW} \left( f \right) = 10^{\alpha}\left(\frac{f}{f_{*}}\right)^{{n_t}_1 } \left[1+\left(\frac{f}{f_{*}}\right)^{{n_t}_2- {n_t}_1}\right]
\end{equation}
evaluated at $\alpha= -9$ at $f_{*}=10^{-2}\ \mathrm{Hz}$, with slopes ${n_t}_1 = 3$ and ${n_t}_2 = -3$. This kind of spectral shape might arise from stochastic GW produced during a first order phase transition around the $\rm{TeV}$ energy scale (see e.g.~refs.\cite{Caprini:2015zlo, Caprini:2019egz} for a dedicated analysis for LISA). The result of the reconstruction is shown in~\cref{fig:bpl}. We have started from 20 initial bins and converged to 8. The reconstructed signal in this case has a large SNR $\simeq  2.1\times 10^4$. Despite the peculiar shape of the signal, the reconstruction is very accurate, also in the external bins, as can be inferred from the small size of the C.L.\ contours, which are shown in~\cref{fig:bplcp}.

\subsection{Bump reconstruction}
\label{sec:bump}
As a last benchmark model we have chosen a \textit{bump} signal characterized by an amplitude $A_*$ and a width $\Delta$ according to
\begin{equation}
     h^2 \Omega_{\rm GW} \left( f \right) =  A_* \, {\rm exp}\left\{ -\frac{[\log_{10}(f/f_*)]^2}{\Delta^2}\right\} \;.
\end{equation}
 This kind of signal is expected from cosmological sources such as non-perturbative effects during post-inflationary preheating, strong first order phase transitions during the thermal era of the universe, or merging of PBH's during the early universe. These phenomena typically generate single- or multi-peaked spectral signals that can be described as a bump. As input parameters for the signal we have chosen $\{A_{0.003}= 10^{-11.0}, \Delta = 0.2\}$.

\begin{figure}[t!]
    \centering
    \includegraphics[width = 0.8 \textwidth]{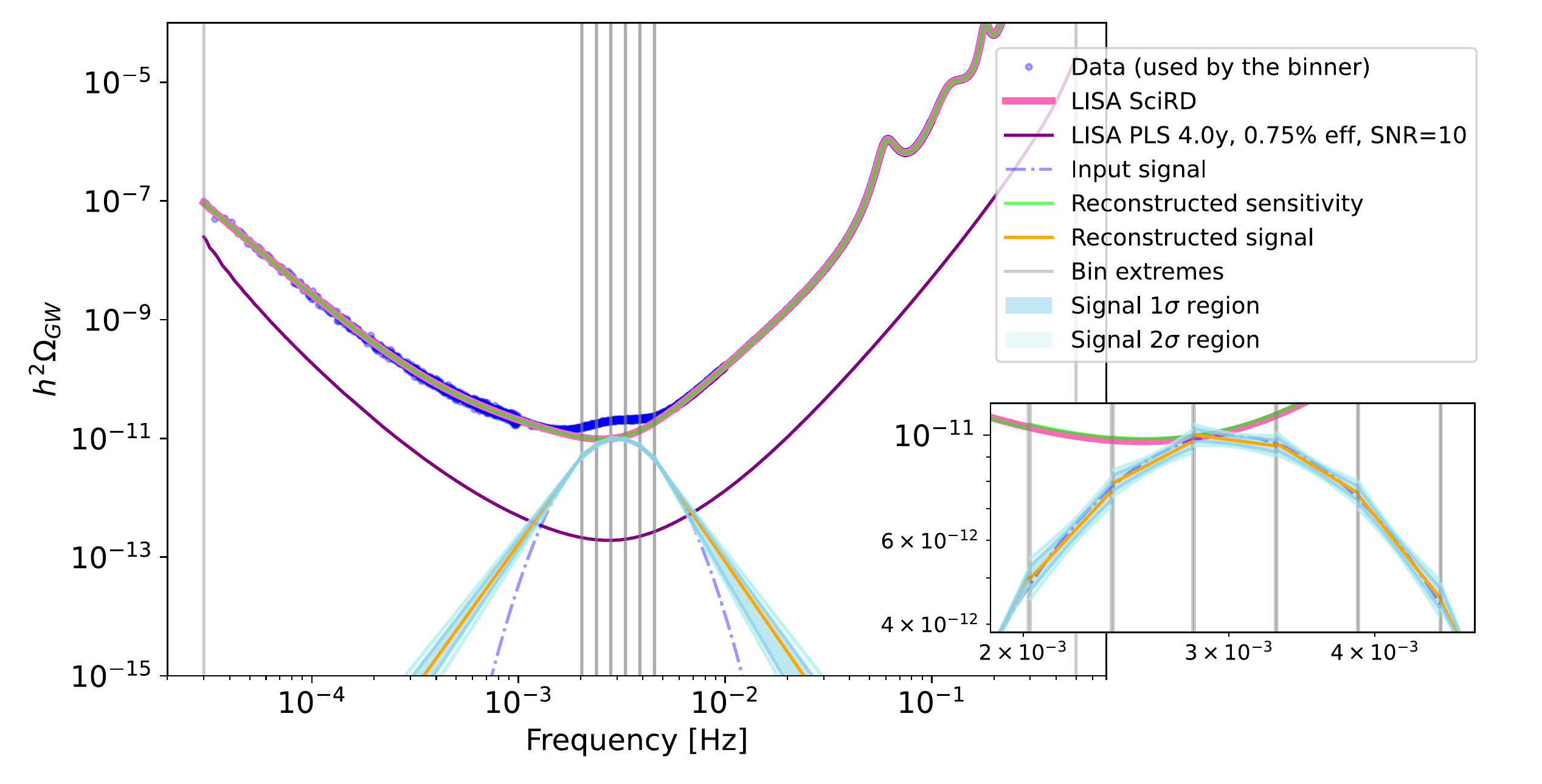}
    \caption{Reconstruction of the bump signal described in~\cref{sec:bump} in the AET channels using 60 initial bins, which converged to 7. In the zoomed regions we can see how the reconstruction accommodates the injected signal within the 1-$\sigma$ band.}
    \label{fig:bumpaet}
\end{figure}

The result of the reconstruction procedure is shown in~\cref{fig:bumpaet}. We have produced it choosing 60 initial bins, which has converged to 7. The reconstruction is more accurate in the central part of the frequency band, where the signal peaks, and worsens in the outer bins. The SNR of the reconstructed bump signal is approximately $339$.

\subsection{Degenerate case}
\begin{figure}
    \centering
    \includegraphics[width = \textwidth]{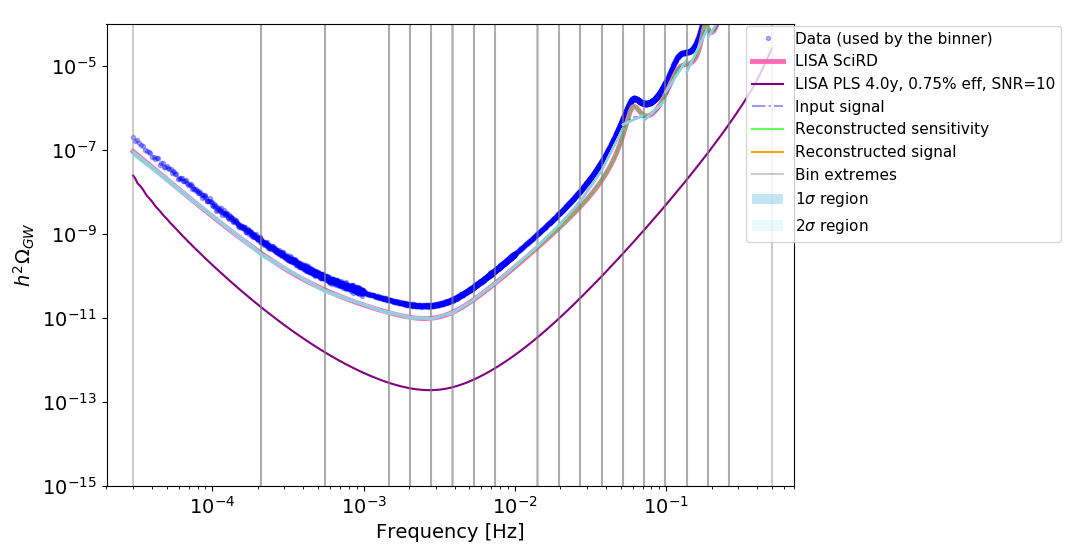}
    \caption{Reconstruction of a signal which is degenerate with noise in XX. Breaking the degeneracy between signal and noise in this case is only possible in a 3-channel analysis (see main text).}
    \label{fig:degenerate}
\end{figure}

One might naively expect that the uncertainties of the parameter reconstruction scale roughly with the square root of the number of independent channels we consider. This guess underestimates the gain in constraining power obtained from a multiple-channel analysis. A key advantage of our three-channel analysis is indeed its capability to break degeneracies. Figure~\ref{fig:degenerate} clearly illustrates this aspect. It considers the (peculiar but didactic) scenario in which in a given channel, say X, the sum of the injected signal and noise matches the noise curve with some acceleration and interferometry-metrology-system parameters larger than the nominal ones. Had we exclusively analysed the XX data, the separation between the signal and the noise  would have been completely dependent on our prior on the noise parameters. In addition, despite the huge SNR (calculated with respect to the nominal sensitivity curve), the errors on the parameters would have been large. On the contrary, thanks to the difference between response functions in the different channel combinations of the AET basis, no signal can be fully degenerate with the noise curve in every channel. The three-channel analysis exploits this feature, and, as can be seen in~\cref{fig:degenerate}, the degeneracy is broken and the reconstruction is reasonably satisfactory.

\subsection{Comparison with and without final MC sampling step}\label{sec:mccomp}
As explained in \cref{sec:sgwbinner_algorithm}, in this work we implement a full-posterior MC as a final step of the algorithm, once the number of bins has been fixed. We expect this step to produce more accurate uncertainty bands for the reconstruction, since, contrary to the main algorithm, it fits to all bins a common noise model.

\begin{figure}
    \centering
    \includegraphics[width=0.8\textwidth]{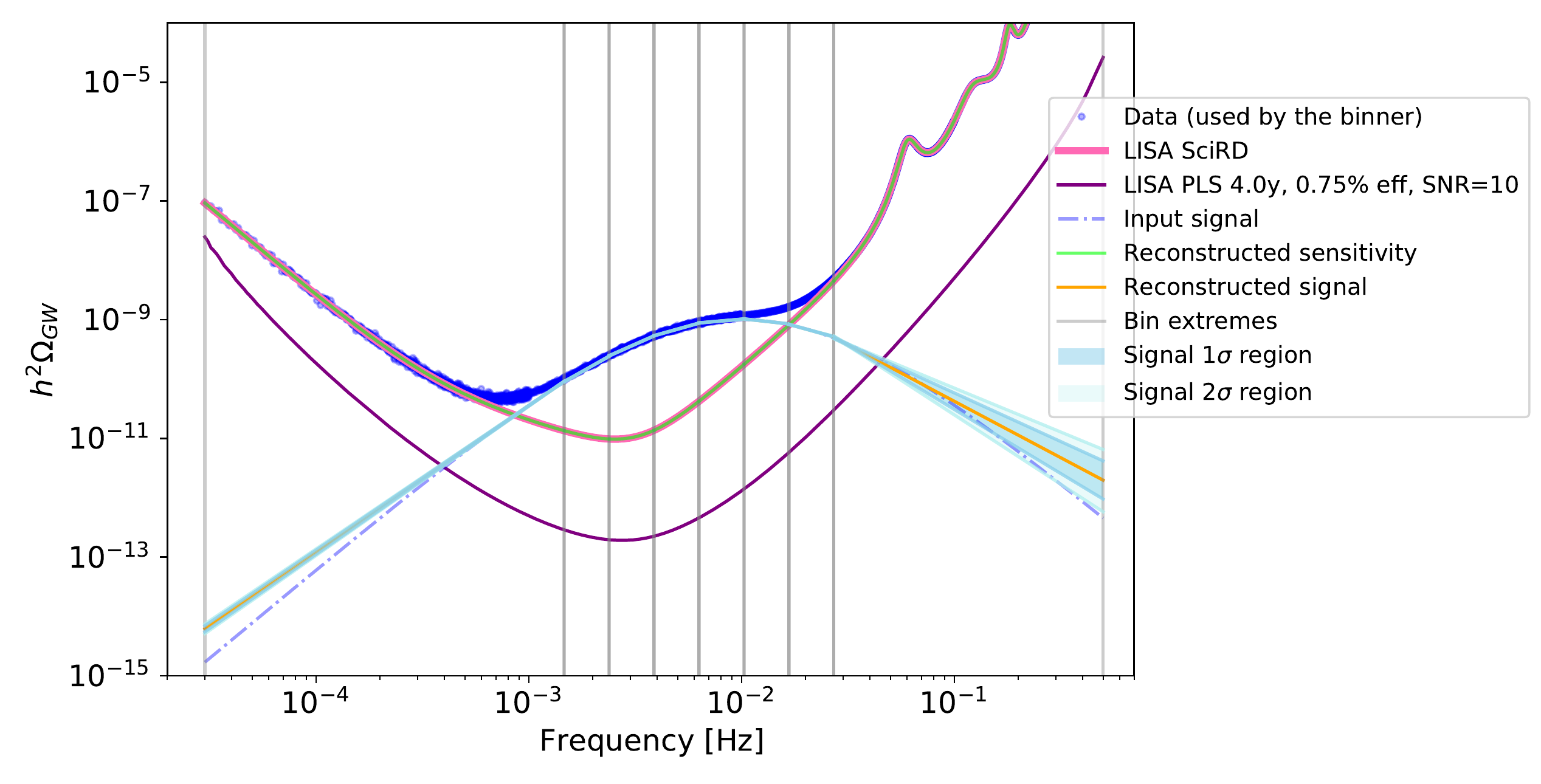}
    \caption{Alternative reconstruction of the broken power law in \cref{fig:bpl}, sampling from the full posterior after having fixed the number of bins. The main effect of considering the full posterior is the 1- and 2-$\sigma$ bands in the outer bins being slightly thinner, due to the small increase in constraining power given by fitting a single noise model to all bins.}
    \label{fig:bpl_mcmc}
\end{figure}

In practice, we found the effect of the per-bin fit of the noise model to affect the reconstruction uncertainties only mildly, at least for moderately high signal-to-noise SGWB shapes (the only ones for which a spectrum reconstruction is expected to produce significant results versus a simple model fit). This is possibly due to the null effect of the noise in the central, high signal-to-noise bins, and to the expected lack of correlation between the left- and rightmost bins, where the noise is the dominant contribution, since the noise in each of these outer bins is affected by only one of the noise parameters: $A$ for the leftmost, low-frequency one, and $P$ for the rightmost, high-frequency one. As en example, in \cref{fig:bpl_mcmc} we show the equivalent result to \cref{fig:bpl}. In this case, as expected, the noise paramters are recovered at 1-$\sigma$.

\section{SGWB reconstruction in the presence of foregrounds}
\label{sec:foreground}
In this section we study the capability of LISA to disentangle different components contributing to the SGWB. In particular, we show that the techniques for the model-independent reconstruction described in the previous sections can be extended to the case in which the signal is a superposition of at least two components, one of which has no reliable model. For this purpose, we consider two mock cases in which the SGWB observed by LISA is the sum of an isotropic astrophysical component (in the following \textit{foreground}) and an unexpected cosmological component. While the latter has no signal model, the spectral shape of the former is assumed to be well known.\footnote{The current models of the foregrounds for LISA still suffer of large uncertainties. It is however reasonable to expect that the foreground models will be quite precise at the end of the LISA mission when, among other things, the population of the sources will be bound by the measurements of the resolvable events.}  We consider two types of foregrounds, each of them expected to capture the qualitative features of some foregrounds really foreseen in LISA:
\begin{description}
    \item[Extra-galactic foreground:] The incoherent superposition of all the extra-galactic compact object mergers that LISA will be unable to resolve, introduce a foreground in the LISA data. Stellar-origin black hole and neutron star binaries should be the main contributors to this extra-galactic astrophysical signal~\cite{Sesana:2016ljz}. For a LIGO-like sensitivity, the foreground of these binaries has a power spectrum (in units of energy density $\Omega_{\rm GW}$) behaving like a power law with power index 2/3 and amplitude $\Omega_{\rm GW} =  8.9^{+12.6}_{-5.6}  \times 10^{-10} $ at $f = 25$\,Hz \cite{LIGOScientific:2019vic}. Extrapolating this information from the LIGO to the LISA frequency band leads to the foreground model~\cite{Sesana:2016ljz, LIGOScientific:2019vic}~\footnote{Several subtleties will have to be addressed to make this model more accurate. It is for instance not obvious that the different resolution power of LISA and LIGO play no relevant role, or that the binary eccentricity evolution is a minor effect.}   
    \begin{equation}\label{eq:ext}
        h^2 \Omega_{\rm GW}  = 10^{\alpha_{FG} } \left( \frac{f}{f_*}\right)^{2/3} \; .
    \end{equation}
    From the mean of the above constraint at LIGO frequencies, we obtain $\alpha_{FG}^* \simeq -12.29$ at $f_*=10^{-3}$Hz, while $\alpha_{FG}^-\simeq-12.72$ from the lower limit. We use these values to build the Gaussian prior $\pi_{FG}(\alpha_{FG}) \sim \mathcal{N}(\alpha_{FG}^*,\sigma^2)$ with $\sigma \simeq 0.43$.
    
    \item[Galactic (averaged) foreground:] The unresolved sub-threshold mergers of galactic binaries constitute a foregrounds with a yearly modulation. A proper treatment of such a signal should take into account the variation in each chunk, and consequently, as shown in ref.~\cite{Adams:2013qma}, the periodicity information can be exploited to partially subtract the foreground. For our illustrative purposes, here we do not follow this approach but consider the signal averaged over the year. This eventually looks like an isotropic and stationary signal with model that in first approximation reads~\cite{Cornish:2017vip, Cornish:2018dyw, Schmitz:2020rag}~\footnote{As noticed in ref.~\cite{Schmitz:2020rag}, the equation for the template in refs.~\cite{Cornish:2017vip, Cornish:2018dyw} contains a typo (a sign flip) in one of the two terms appearing in the exponent.}
    \begin{equation}\label{eq:gal}
        h^2 \Omega_{\rm GW}  = 10^{\alpha_{FG} } f^{2/3} \textrm{e}^{- f^a_1 - a_2 f \sin(a_3 f )   } \left\{  1 + \tanh\left[ a_4 (f_k -f) \right] \right\}
         \; .
    \end{equation}
    For concreteness, we take the values quoted in Table 1 of~ref.~\cite{Cornish:2018dyw} for the parameters $a_1, \dots, a_4, f_k$, choosing four years of  observation time. We assume these parameters to be known so that no fit on them will be performed. The fit is instead carried out over the amplitude parametrised by $\alpha_{FG}$. As a toy example, we  build the prior on $\alpha_{FG}$ following some reasonable but somehow arbitrary criteria. We assume the prior to be Gaussian and, following ref.~\cite{Cornish:2018dyw}, we take it to peak at  $\alpha_{FG}^* = -7.95$. To set the expected deviation interval, we use an uncertainty similar to the one of the sensitivity curve. This roughly corresponds to impose $\alpha_{FG}\in [-8.21, -7.79]$ at $68\%$ C.L.\,. The Gaussian prior then becomes $\pi_{FG}(\alpha_{FG}) \sim \mathcal{N}(\alpha_{FG}^*,\sigma^2)$ with $\sigma \simeq 0.28$.\footnote{Notice that the $20\%$ margins on the two noise parameters induce a $45\%$ variation on the level of the sensitivity curve. The values of $\sigma$ that follows, makes the prior slightly broader than the estimate prediction. This choice is motivated by the large uncertainties in the prior definition, but should not play a big role in this proof-of-concept example.}
\end{description}

As for the cosmological component, we inject a broken power-law signal, as described in~\cref{eq:Broken_pl}, with amplitude $\alpha= -9$ at $f_{*}=10^{-2}\ \mathrm{Hz}$, and slopes ${n_t}_1 = 6$ and ${n_t}_2 = -4$. As stated, the model for the foreground is assumed to be known, so we are not interested in reconstructing its shape but only its amplitude. For this reason, the foreground parameter can in practice be treated as an extra noise parameter. Following the procedure described in~\cref{sec:mccomp}, after we have determined the final number of bins, we perform a final MC sampling in which the foreground amplitude is fit simultaneously for all bins, as the noise parameters are. In the examples below we have used \texttt{PolyChord} \cite{2015MNRAS.450L..61H,2015MNRAS.453.4384H} as the MC sampler, via its interface with \texttt{Cobaya} \cite{Torrado:2020dgo}. The results were analyzed and the contour plots were produced using \texttt{GetDist} \cite{Lewis:2019xzd}.

\begin{figure}
  \begin{subfigure}{0.95\textwidth}
    \centering \includegraphics[width=\textwidth]{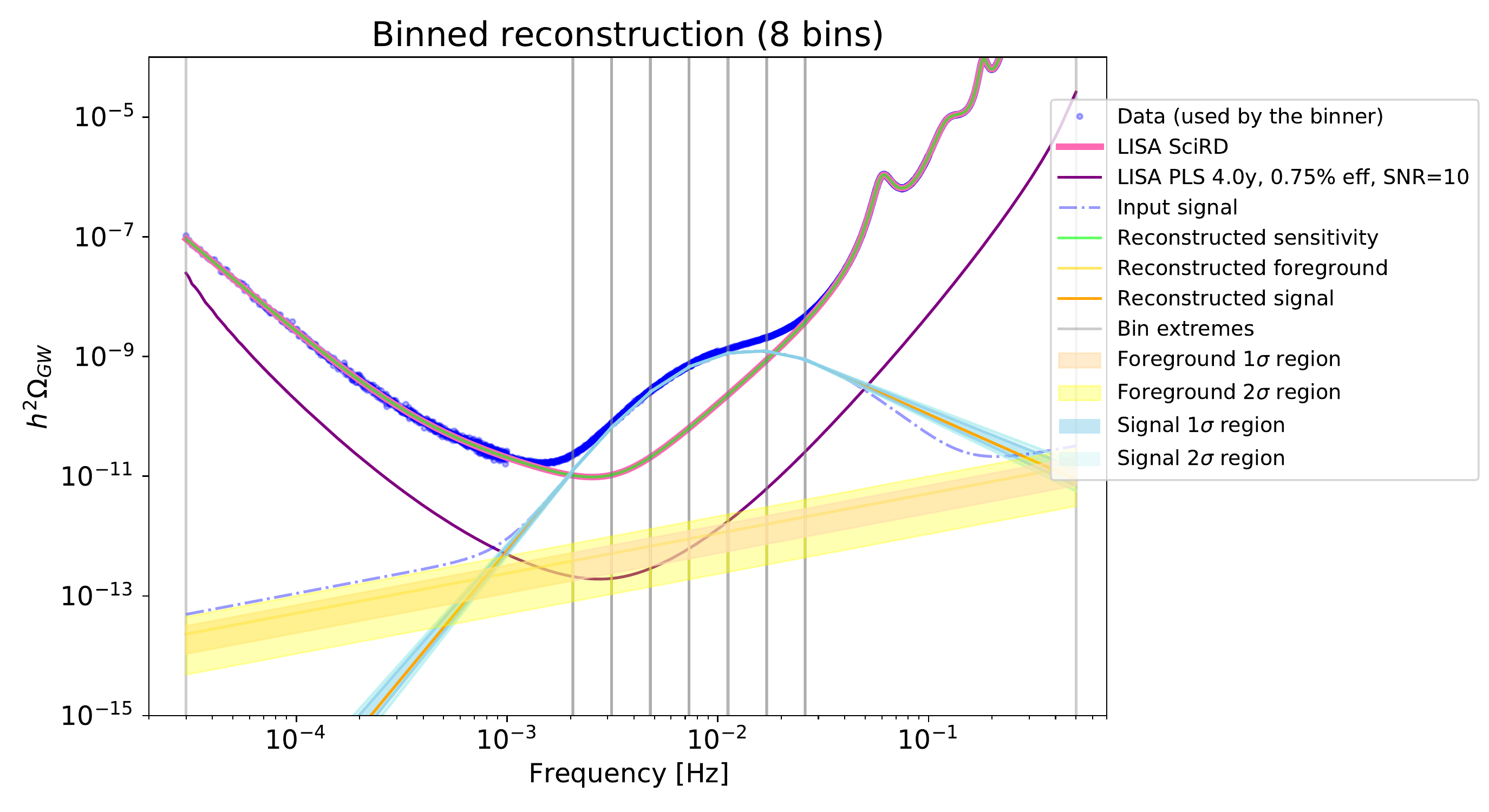}
    \caption{}\label{fig:ext:bands}
  \end{subfigure}
  \begin{subfigure}{0.45\textwidth}
    \centering \includegraphics[width=\textwidth]{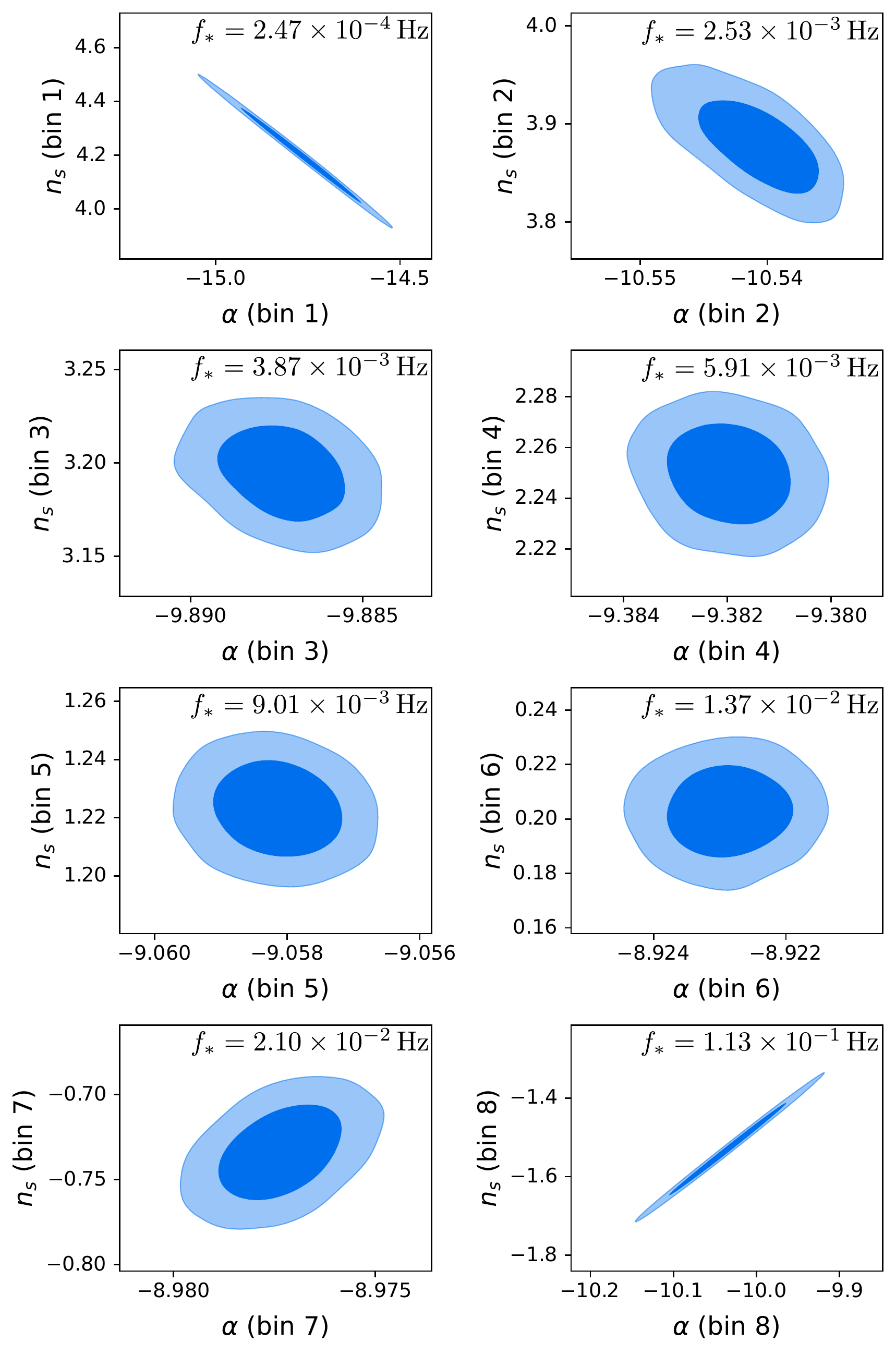}
    \caption{}\label{fig:ext:bins}
  \end{subfigure}
  \begin{subfigure}{0.54\textwidth}
    \centering \includegraphics[width=\textwidth]{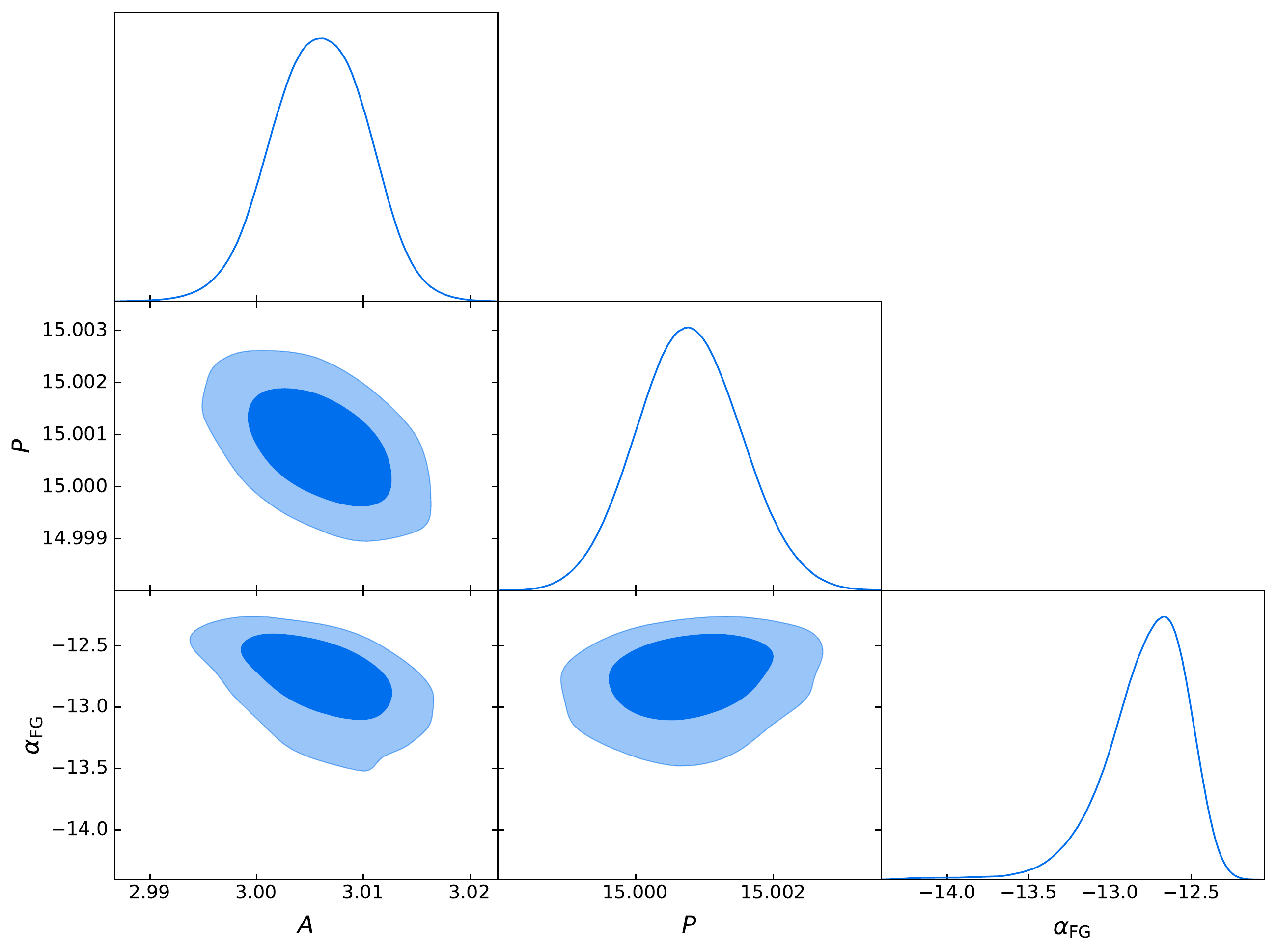}
    \caption{}\label{fig:ext:fgnoise}
  \end{subfigure}
\caption{\textbf{(a)} Simultaneous reconstruction of a broken power-law signal (see~\cref{sec:Broken_PL}), an extra-galactic foreground component modelled as the power law in~\cref{eq:ext}, and the noise spectrum. The initial number of bins was 23. Contour plots for the bin parameters and a triangle plot of the marginalized posterior for the two noise and one foreground parameters are shown respectively in \textbf{(b)} and \textbf{(c)}.}
\label{fig:ext}
\end{figure}

In~\cref{fig:ext} we show the reconstruction of the broken power-law SGWB signal described above in the presence of the extra-galactic foreground component described above. As can be seen in~\cref{fig:ext:bands}, wherever the injected signal is sufficiently large with respect to the LISA sensitivity, the signal reconstruction is practically unaffected by the presence of the foreground. Contour plots for the parameters of the binned model are shown in~\cref{fig:ext:bins}. Except for the two outer bins, where the amplitude and the tilt of the signal are degenerate, the measurements of the log amplitude and of the tilt are consistently accurate. A triangle plot for the two LISA noise parameters together with the log of the foreground amplitude is shown in~\cref{fig:ext:fgnoise}. Only mild degeneracy are found and all the parameters are consistent with the injected values at $2\sigma$. Notice that the measurement of the foreground amplitude does not improve over its prior, since in this example the cosmological signal is much larger than the foreground where the sensitivity is highest. How this may change if we consider different shapes and amplitudes for the cosmological component is beyond the scope of this work, and is left to future studies on the topic.

\begin{figure}
  \begin{subfigure}{0.95\textwidth}
    \centering \includegraphics[width=\textwidth]{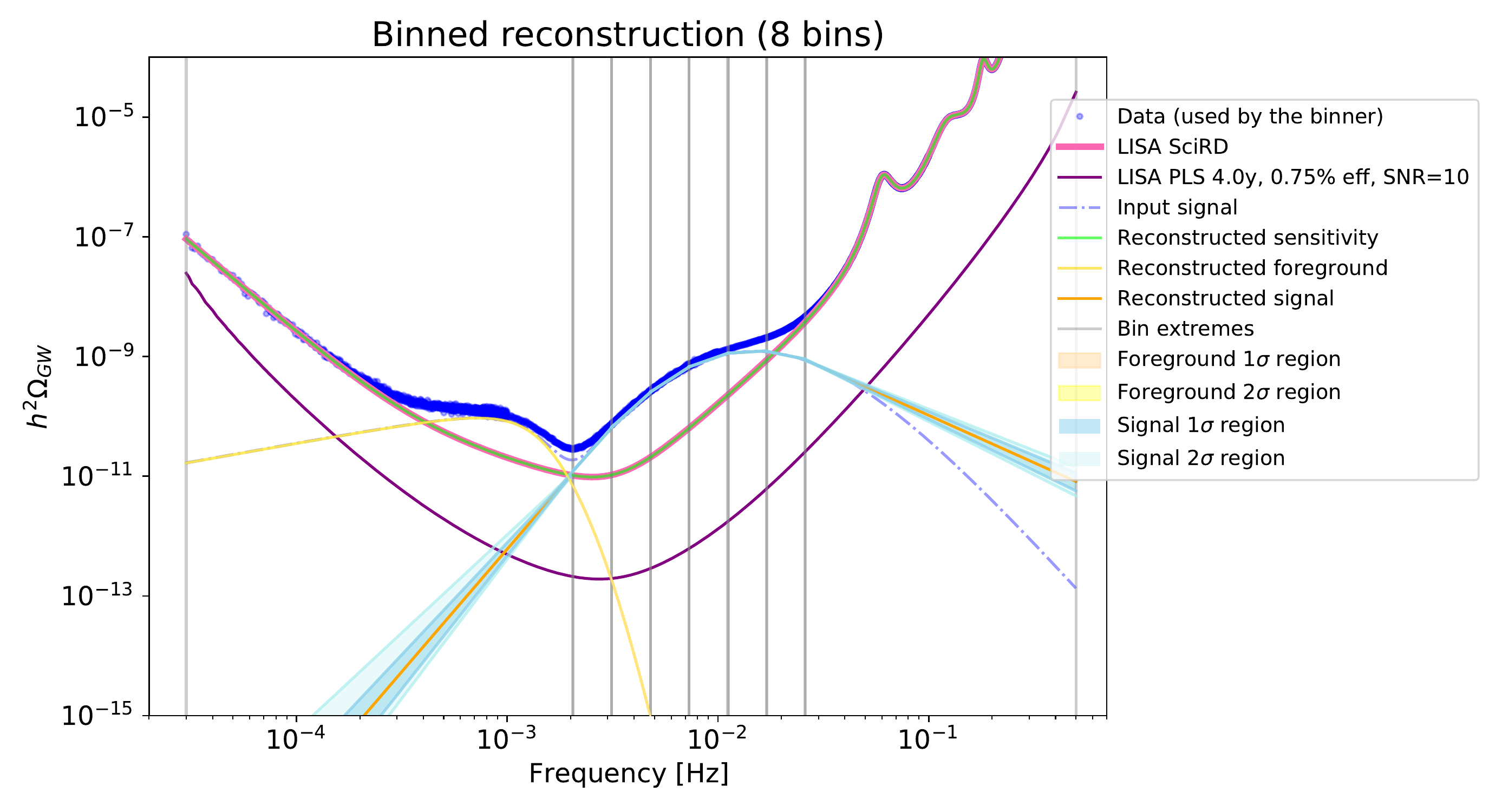}
    \caption{}\label{fig:gal:bands}
  \end{subfigure}
  \begin{subfigure}{0.45\textwidth}
    \centering \includegraphics[width=\textwidth]{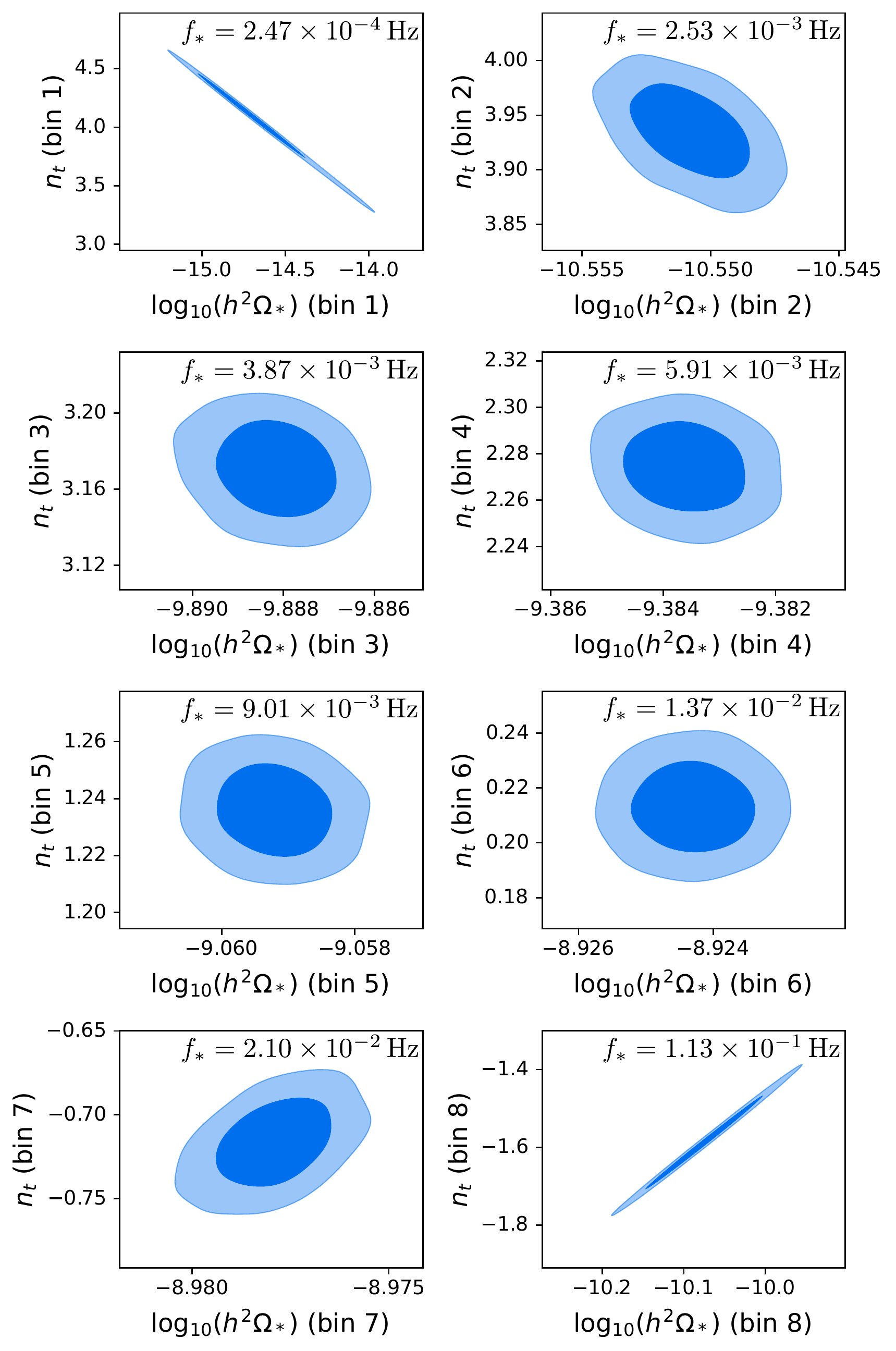}
    \caption{}\label{fig:gal:bins}
  \end{subfigure}
  \begin{subfigure}{0.54\textwidth}
    \centering \includegraphics[width=\textwidth]{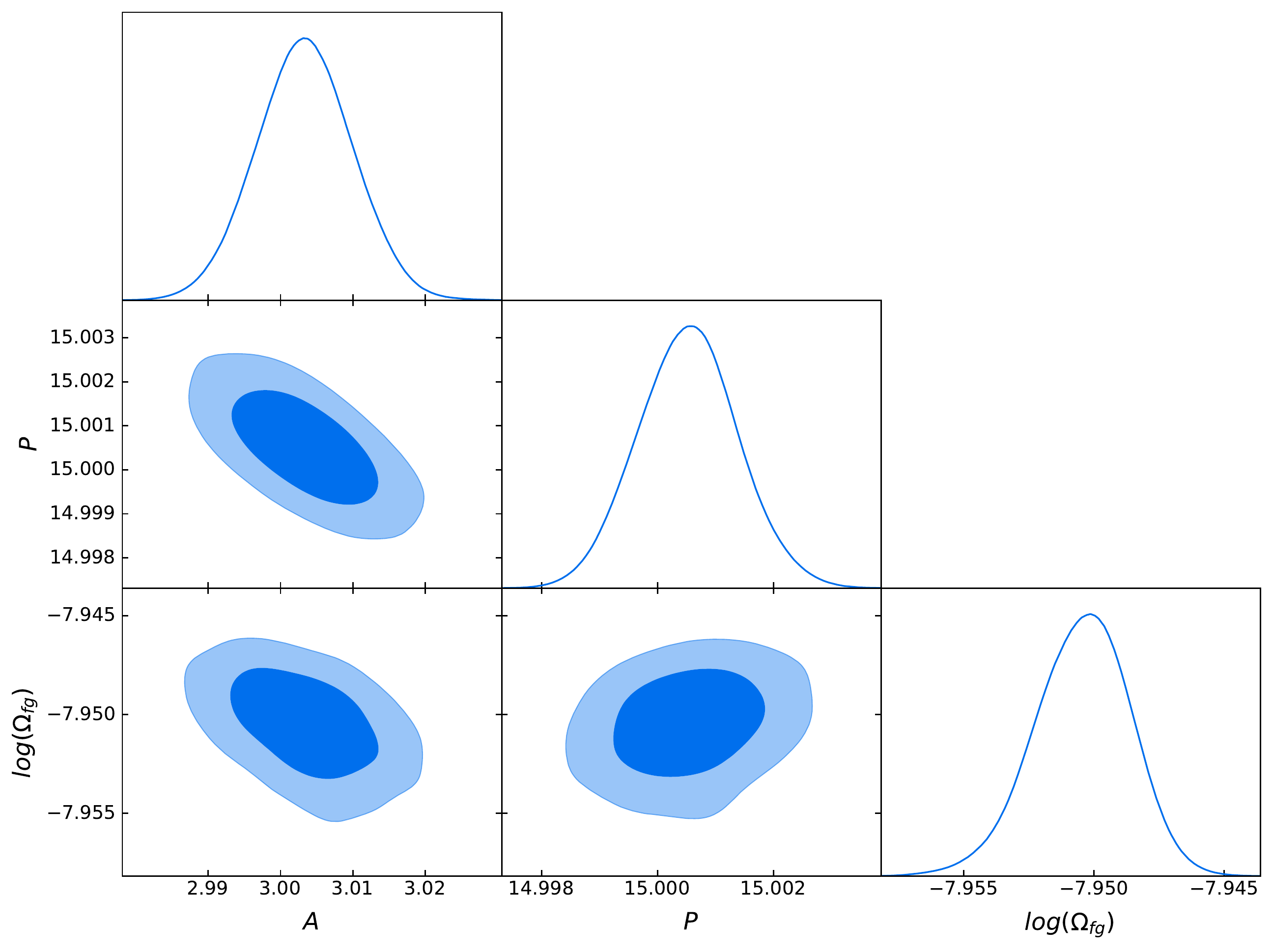}
    \caption{}\label{fig:gal:fgnoise}
  \end{subfigure}
\caption{\textbf{(a)} Simultaneous reconstruction of a broken power-law signal (see~\cref{sec:Broken_PL}), an galactic foreground component with a model given in \cref{eq:gal}, and the noise spectrum. The initial number of bins was 23. Contour plots for the bin parameters and \ a triangle plot of the marginalized posterior for the two noise and one foreground parameters are shown respectively in \textbf{(b)} and \textbf{(c)}.}
\label{fig:gal}
\end{figure}

The reconstruction of the same cosmological signal, now in the presence of the galactic foreground described above, is shown in \cref{fig:gal}. In this case, the cosmological and the foreground SGWB components are the dominant contribution in different segments of the LISA frequency band. This allows both for a very accurate reconstruction of the cosmological component, and a precise measurement of the amplitude of the foreground component (well beyond its prior), as well as of the noise parameters. This is clear from the contour plots and from the triangle plot in~\cref{fig:gal:bins,fig:gal:fgnoise}.

\section{ Conclusions}
\label{sec:conclusions}
In this paper we have extended the work initiated in ref.~\cite{Caprini:2019pxz} from a single channel to the use of the three TDI 1.5 channels for the LISA space-based interferometer. We have presented a formalism and an algorithm to extract a SGWB signal from the LISA data, assuming a simplified model for the noise power spectral density and two illustrative foregrounds. We have worked in the AET channel combination, leading to vanishing off-diagonal terms in channel space when the LISA's arms have equal length, and the noise power spectral densities of all test masses are the same. Under these idealised conditions, we are able to simplify the numerical computations by converting to the noise orthogonal AET TDI variables. In addition, we find that under these assumptions, the results are identical to using the standard XYZ TDI variables.

While in the code we have worked with the numerical response functions, we have provided the analytic form for the XX channel (as well as the YY and ZZ channels), and for the $\text{AA}$ (as well as the $\text{EE}$) and for the $\text{TT}$ channels. We found that in the TT channel both the signal response and the acceleration noise power spectrum are suppressed up to $\mathcal{O}(10^{-2})$ Hz. As a consequence, TT can be used to impose a strong constraint on the interferometric component of the noise, and a weaker but substantial one on the acceleration noise component, thus making possible to use the full frequency range for the reconstruction, which was not possible in our previous single-channel approach. For the same reason, only two channels contribute to the constraints on the SGWB signal, which are thus improved by $\sqrt{2}$ with respect to the single-channel case.

We have updated the pipeline presented in ref.~\cite{Caprini:2019pxz} to simulate data in frequency space to the TDI 1.5 three-channel case, and constructed a likelihood that produces unbiased constraints on the signal and the noise in the low SNR regime (if a signal were clearly detected in the future, the pipeline would have to be updated accordingly). After having established the formalism and updated our \texttt{SGWBinner} algorithm accordingly, we have applied it on three physically-motivated scenarios (i.e. a power-law, a broken power-law and a bump GW signal), where we find it to be able to reconstruct the injected signals accurately. We check the per-bin optimisation approach of the \texttt{SGWBinner} against an MC exploration of the full posterior and we found a good matching of the results. Finally, in order to test the robustness of our reconstruction algorithm in presence of backgrounds and (astrophysical) foregrounds, we have performed some tests on signals given by the sum of an isotropic astrophysical component and a (unexpected) cosmological component.  As astrophysical foregrounds we have considered an extra-galactic-like component and a galactic-like component, which are both foreseen in the LISA band. 
Besides being a robustness check, such a test is also relevant for testing disentangling ability for different signal contributions.
We have seen that, when the injected signal is sufficiently large with respect to the LISA sensitivity, the signal reconstruction is practically unaffected by the presence of the foregrounds.

In future work, we plan to apply our methodology to more realistic scenarios, both from an instrumental and an astrophysical point of view. The performance of our techniques will be tested against data sets that contain spurious and confusion signals as well, such as the compact binaries in the vicinity of our galaxy~\cite{Korol:2018wep}. Moreover, the case of varying constellation arm lengths, and unequal power spectra densities of the test-masses noises will be tested. This would further increase the complexity of the analysis by adding extra degrees of freedom and degrade the constraining capacity in our reconstruction. Nevertheless we expect the simplified configuration studied here to be a reasonable approximation for exploratory studies of SGWB.

\acknowledgments
We thank Kai Schmitz for bringing to our attention some typos in the literature of the galactic foregrounds, and Marco Peloso for noticing inconsistent conventions in the appendix. We acknowledge the LISA Publication and Presentation committee, in particular Nelson Christensen for useful comments. R.F.~was supported in part by the Department of Energy under Grant No.~DE-SC0009919, and a grant from the Simons Foundation/SFARI 560536. The work of M.P.~was supported by Science and Technology Facilities Council consolidated grant ST/P000762/1. M.P.~was supported in part by the National Science Foundation under Grant No.NSF PHY-1748958. R.F.~and M.P.~would like to thank the Kavli Institute for Theoretical Physics at UC Santa Barbara for the kind hospitality during part of this work. G.N.~is partly supported by the ROMFORSK grant Project.~No.~302640, and acknowledges networking support by the GWverse COST Action CA16104, “Black holes, gravitational waves and fundamental physics.". N.K.~was supported by the CNES DIA-PF post-doctoral fellowship program. A.R.~acknowledges funding from Italian Ministry of Education, University and Research (MIUR) through the ``Dipartimenti di
eccellenza'' project Science of the Universe. 

\appendix
    
\section{Signal and detector technicalities}
\label{sec:technicalities}
\subsection{Polarization tensors}
\label{sec:formulae}
We work in natural units and in the Lorentz transverse-traceless gauge. We consider the coordinate system $\{\hat e_x, \hat e_y, \hat e_z\}$ arbitrarily oriented and at rest with respect to the isotropic SGWB we analyse. We neglect the motion of LISA in this frame; taking it into account is conceptually straightforward~\cite{Domcke:2019zls} but adds some  technical complications that are unnecessary for the purpose of the present paper.
In this reference frame, the $\mathbf{k} $ wave-vector of an incoming plane GW sets the orthonormal basis~\cite{Bartolo:2018qqn,Domcke:2019zls}
\begin{equation}
\hat{u}(\hat{k}) = \frac{\hat{k} \times \hat{e}_z}{|\hat{k} \times \hat{e}_z|} \; , \hspace{4cm} \hat{v} (\hat{k} )= \hat{k} \times \hat{u} \;, 
\end{equation}
(where $\hat k$ is unit vector in the direction of $\mathbf{k}$ and we will denote its magnitude by $k=|\mathbf{k}|$) that we introduce to define the ``plus" ($+$) and ``cross" ($\times$) polarization tensors:
\begin{equation}
\label{eq:PlusCrossTensors}
e_{ab}^{(+)} (\hat{k})= \frac{{\hat u}_a {\hat u}_b - {\hat v}_a {\hat v}_b }{\sqrt{2}}  \; , \hspace{4cm} 
e_{ab}^{(\times)}(\hat{k}) = \frac{{\hat u}_a {\hat v}_b + {\hat v}_a {\hat u}_b }{\sqrt{2}} \;. 
\end{equation} 
Since $\hat{u} (- \hat{k} ) = - \hat{u} (\hat{k} )$ and  
$\hat{v} (- \hat{k} ) = \hat{v} (\hat{k})$, 
the tensors $e_{ab}^{(+)}$ and $
e_{ab}^{(\times)}$ fulfill the conditions
\begin{equation}
\begin{aligned}
& e^{+/\times}_{ab} (\hat{k} )  = e^{+/\times \, *}_{ab} (\hat{k})  \;,  \hspace{1cm} 
&& e^{+}_{ab} (\hat{k} )   = e^{+}_{ab} (- \hat{k} )   \;,  \hspace{1cm} 
&& e^{\times}_{ab} (- \hat{k} )  = - e^{\times}_{ab} (\hat{k} ) \; ,\\
& e^+_{ab} (\hat{k})  e^+_{ab}(\hat{k}) = 1 \; , 
&& e^\times_{ab} (\hat{k}) e^\times_{ab}(\hat{k})  = 1 \;, &&  e^+_{ab} (\hat{k}) e^\times_{ab} (\hat{k})  = 0 \; .
\end{aligned}
\end{equation}
Analogously, we introduce the ``right-handed" (R) and ``left-handed" (L) polarization tensors:
\begin{equation}
e_{ab}^{R} (\hat{k} ) = 
\frac{ {\hat u}_a  + i \,  {\hat v}_a  }{\sqrt{2}} \, 
\frac{ {\hat u}_b  + i \,  {\hat v}_b  }{\sqrt{2}} 
\;, \hspace{0.5cm} 
e_{ab}^{L} (\hat{k} ) = 
 \frac{ {\hat u}_a  - i \,  {\hat v}_a  }{\sqrt{2}} \, 
\frac{ {\hat u}_b  - i \,  {\hat v}_b  }{\sqrt{2}} 
\ \;.  
\label{u-v-oneindex} 
\end{equation} 
They are related to the plus and cross polarization basis by means of the relationships
\begin{eqnarray}\label{eq:ChiralTensors}
e_{ab}^{R} (\hat{k} ) = \frac{e_{ab}^{+} + i \, e_{ab}^{\times}}{\sqrt{2}}  \;, \hspace{1cm} 
e_{ab}^{L} (\hat{k} ) = \frac{e_{ab}^{+} - i \, e_{ab}^{\times}}{\sqrt{2}}  \;.
\end{eqnarray}

The superposition of all GWs reaching the position $x$ at the time $t$ can be expressed in terms of incoming plane waves and reads 
\begin{equation}
\label{eq:h_of_t_k}
h_{ab}(\mathbf{x},t)  = \int_{- \infty}^{+ \infty} \textrm{d} f \int_{\Omega} \textrm{d} \Omega_{\hat{k} }\; \textrm{e}^{2\pi i f (t - \hat{k} \cdot \mathbf{x})}  \; \sum_{A} \tilde{h}_A (f, \hat{k})  \; e_{ab}^{A}(\hat{k}) \;  ,
\end{equation}
where $f$ is the frequency of each plane wave, $\textrm{d} \Omega_{\hat{k}}$ is the infinitesimal solid angle from which the incoming wave with wave-vector $\mathbf{k}$ arrives, and finally $\tilde{h}_A (f, \hat{k}) \equiv  k^2 \, \tilde{h}_A (\mathbf{k}) $. The index $A$ can be used to label either the plus and cross polarizations or the left- and right-handed polarizations, so that the sum can run indifferently over one or the other basis (i.e.~$A=+,\times$ or $A=L,R$). So that finally we can write
\begin{equation}
h_{ab}(\mathbf{x},t)  =   \int \textrm{d}^3 k \; \textrm{e}^{- 2\pi i \mathbf{k} \cdot \mathbf{x} } \sum_{A}  \left[  \textrm{e}^{2\pi i k t }  \; \tilde{h}_A (\mathbf{k})  \; e_{ab}^{A}(\hat{k})  +  \textrm{e}^{-2\pi i k t }  \; \tilde{h}_A^* (-\mathbf{k})  \; e_{ab}^{A \, *}(-\hat{k})\right] \; .
\end{equation}

\subsection{TDI variables}
\label{sec:TDIvariables}

Consider two test masses inside two of the three LISA spacecrafts. These test masses, labelled 1 and 2, are located at $\mathbf{ x}_1$ and $\mathbf{x}_2$ and separated by the vector $L \hat{l}_{12}$, with $ \hat{l}_{12}= (\mathbf{x}_2 - \mathbf{x}_1 )/ |\mathbf{ x}_2 - \mathbf{x}_1 |$ and  $L=2.5\times 10^9\,$m. When a GW crosses the detector, a photon leaving test mass $2$ at time $t-L$ is received at test mass 1 with the time shift~\cite{estabrook, Romano:2016dpx}
\begin{equation}
	\Delta T_{12}(t) = \frac{\hat{l}^a_{12} \hat{l}^b_{12} }{2 } \int_{0}^{L} \textrm{d}s \, h_{ab} (t(s), \mathbf{x}(s) ) \; .
\end{equation}
For the photon path, the leading order approximations $t(s)=t-L+s$ and $\mathbf{ x}(s)= \mathbf{ x}_2 -s \hat{l}_{12} $ suffice. It follows
\begin{equation}
\begin{aligned}
	\Delta T_{12}(t) =  L \int \textrm{d}^3 k \; \textrm{e}^{ -2\pi i \mathbf{k} \cdot \mathbf{x}_2 }  \sum_{A} 
	&
	\left[ \textrm{e}^{2\pi i k (t-L) } 	\mathcal{M} (\mathbf{k}, \hat{l}_{ij})  \; \tilde{h}_{A } (\mathbf{k})  \; \mathcal{G}^{A }(\hat{k}, \hat{l}_{12})  \right. \\
	 &
	+ \left. \textrm{e}^{-2\pi i k (t-L) } 	\mathcal{M}^* (-\mathbf{k}, \hat{l}_{ij})  \;  \tilde{h}^*_A (-\mathbf{k})  \; \mathcal{G}^{A \ *}(-\hat{k}, \hat{l}_{12}) \right]  \; ,
\end{aligned}
\end{equation}
with
\begin{equation}
	\mathcal{G}_i^A(\hat{k}, \hat{l}_{ij})  \equiv  \frac{\hat{l}^a_{ij} \hat{l}^b_{ij} }{2} \, e_{ab}^{A}(\hat{k})  \; , \qquad 
		\mathcal{M} (\mathbf{k}, \hat{l}_{ij}) \equiv \textrm{e}^{\pi i k L (1 + \hat{k} \cdot \hat{l}_{ij} ) }  \text{sinc}\left[ kL  (1 + \hat{k} \cdot \hat{l}_{ij} ) \right] \; ,
\end{equation}
and $\text{sinc} (x) \equiv 	\sin(\pi x) /(\pi x)$.

In practice, LISA does not observe this time shift, but instead observes the corresponding Doppler effect. Let us assume that the test mass 2 emits photons with frequency $\nu$.  The test mass 1 receives such photons with the fractional Doppler frequency shift $\Delta F_{12} (t) \equiv 
 \Delta \nu_{12} (t)/\nu = -
 \textrm{d} \Delta T_{12} (t)/\textrm{d} t $.

In LISA neither the frequency $\nu$ nor the positions of the test masses are not perfectly known. For this reason, the observation of a GW requires to perform time delay interferometry (TDI), that is, to compare the fractional frequency shifts of the same light pulse split into different paths. Moreover, to take into account the uncertainties on the positions of the test masses, the relative frequency shifts are measured 
after conveying the laser pulses along paths that are quite complex. Without reaching the level of complexity of the paths adopted in LISA, which are still under discussion, hereafter we focus on paths that contain all the conceptual elements of the  realistic ones. 

A key element in common to every TDI measurement is the fractional Doppler shift of a photon following paths such that photon frequency noise and motion of optical benches cancels. For concreteness let us consider the case of the closed path $1 \to 2 \to 1$, with a photon starting from $\mathbf{x}_1$ at  time $t-2L$ and returning back to $\mathbf{x}_1$ at time $t$. In this case the final fractional Doppler shift is 
\begin{equation}
\begin{aligned}
\Delta F_{1(2)} (t) & \equiv \Delta F_{21}(t -L) +  \Delta F_{12}(t) \\
 & = - \int \textrm{d}^3 k \; \textrm{e}^{ -2\pi i \mathbf{k} \cdot \mathbf{x}_1 }  \frac{i k}{f_\star}
\sum_{A}  \left[ \textrm{e}^{2\pi i k (t -L) }  \mathcal{T}(\mathbf{k}, \hat{l}_{12}) \; \tilde{h}_{A } (\mathbf{k})  \; \mathcal{G}^{A }(\hat{k}, \hat{l}_{12})  \right. \\ 
& \hspace{4cm}- \left. \textrm{e}^{-2\pi i k (t -L) }  \mathcal{T}(\mathbf{k}, \hat{l}_{12}) \;  \tilde{h}^*_A (-\mathbf{k})  \; \mathcal{G}^{A \ *}(-\hat{k}, \hat{l}_{12}) \right] \; ,
\end{aligned}
\label{eq:DeltaF12}
\end{equation}
where $f_\star =(2\pi L)^{-1}$ is the detector characteristic frequency, and $\mathcal{T}$ is the so called transfer function defined as
\begin{equation}
\label{eq:transfer_function}
	\mathcal{T} (\mathbf{k}, \hat{l}_{ij})  \equiv \textrm{e}^{\pi i k L (1 - \hat{k} \cdot \hat{l}_{ij} ) }  \text{sinc}\left[ kL  (1 + \hat{k} \cdot \hat{l}_{ij} ) \right] + \textrm{e}^{- \pi i k L (1 + \hat{k} \cdot \hat{l}_{ij} ) }  \text{sinc}\left[ kL  (1 - \hat{k} \cdot \hat{l}_{ij} ) \right] \; .
\end{equation}
We can now compare the fractional Doppler shifts of two laser beams of identical frequency, with the first one following the path $1 \to 2 \to 1$ and the second following the path $1 \to 3 \to 1$. Their relative Doppler shift measured at 1 at the time $t$, here denoted with $\Delta F_{1(23)}(t)$, results
\begin{equation}
	  \Delta F_{1(23)}(t) = \Delta F_{1(2)}(t)-\Delta F_{1(3)}(t)  \; .
	  \label{eq:tdi1}
\end{equation}
Closer inspection reveals that this variable is only a good TDI variable for an equilateral configuration. For configurations that depart from the equilateral configuration more complicated TDI variables are required. For a configurations that depart from an equilateral configuration, but for which the time variation of the armlengths is slow, a path $1\to 2\to 1 \to 3 \to 1$ and the other following the path $1\to 3\to 1 \to 2 \to 1$ lead to the relative fractional Doppler shift
\begin{equation}
	    \Delta F^{1.5}_{1(23)}(t) = \Delta F_{1(23)}(t -2L) + \Delta F_{1(32)}(t)  
	     \qquad     \qquad \textrm {(TDI 1.5 variable)} \; .
	     \label{eq:tdi15}
	\end{equation}
We will work with these TDI 1.5 variables and introduce the shorthand notation $\text{X}=\Delta F^{1.5}_{1(23)}$. The TDI variables $\text{Y}$ and $\text{Z}$ are obtained by cyclic permutation.  
However, our techniques can easily be adapted to more complex TDI variables that also account for the time evolution of the armlengths.

\subsection{Response functions}
\label{sec:response}
By substituting eq.~\eqref{eq:DeltaF12}
into eq.~\eqref{eq:tdi1}, one can see that the signal contribution to $\Delta F_{1(23)}(t)$ is 
\begin{equation}
\begin{aligned}
\Delta F_{1(23)}(t)	&  =  \int \textrm{d}^3 k \; \textrm{e}^{ -2\pi i \mathbf{k} \cdot \mathbf{x}_1 } (-2\pi i k L) \sum_{A} \left[  \textrm{e}^{2\pi i k (t -L) } R_1^A(\mathbf{k}, \hat{l}_{12}, \hat{l}_{13} )  \tilde{h}_{A } (\mathbf{k})  +  \right. \\ 
 & \hspace{4cm}- \left. \textrm{e}^{-2\pi i k (t -L) } {R_1^{A}}^*(-\mathbf{k}, \hat{l}_{12}, \hat{l}_{13} )  \tilde{h}_{A}^* (-\mathbf{k})\right] \; , 
\end{aligned} 
\end{equation}
with
\begin{equation}
\label{eq:R_func}
R_i^A(\mathbf{k}, \hat{l}_{ij}, \hat{l}_{ik} ) \equiv   \mathcal{G}^{A }(\hat{k}, \hat{l}_{ij})  \mathcal{T}(\mathbf{k}, \hat{l}_{ij}) - \mathcal{G}^{A }(\hat{k}, \hat{l}_{ik})  \mathcal{T}(\mathbf{k}, \hat{l}_{ik})  \; .
\end{equation}
The functions $R_i^A$ are the so called LISA response function, and describe how the detector responds to a plane wave of wave-vector $\mathbf{ k}$ when the LISA test masses $i$ and $j$ are oriented along the direction $\hat{l}_{ij}$. It then immediately follows that
\begin{equation}
\begin{aligned}
\Delta F^{1.5}_{1(23)}(t)	&  =  \int \textrm{d}^3 k \; \textrm{e}^{ -2\pi i \mathbf{k} \cdot \mathbf{x}_1 } (-2\pi i k L) \sum_{A} \left[  \textrm{e}^{2\pi i k (t -L) } W(kL) R_1^A(\mathbf{k}, \hat{l}_{12}, \hat{l}_{13} )  \tilde{h}_{A } (\mathbf{k})  +  \right. \\ 
 & \hspace{4.8cm}- \left. \textrm{e}^{-2\pi i k (t -L) }W^*(kL) {R_1^{A}}^*(-\mathbf{k}, \hat{l}_{12}, \hat{l}_{13} )  \tilde{h}_{A}^* (-\mathbf{k})\right] \; , 
 \nonumber
\end{aligned} 
\end{equation}
with $W(kL) \equiv  \textrm{e}^{- 4\pi i k  L } -1 $. Notice that $
  |	W(kL) |^2 = 2 \left[ 1 - \cos(4\pi k  L) \right] = 4 \sin^2(2\pi k  L)$.
The two-point correlation functions of  $\Delta F_{i(jk)}(t)$ can then be expressed as
\begin{equation}
\label{eq:two_point_dfreq}
\begin{aligned}
\left\langle \Delta F_{i(jk)}(t) \Delta F_{l(mn)}(t)  \right\rangle & =   \int \textrm{d}^3 k  \; \textrm{e}^{ -2\pi i  \mathbf{k} \cdot (\mathbf{x}_i - \mathbf{x}_l) } 2(2\pi k L)^2 |	W(kL) |^2 \times  \\
& \hspace{1cm} \times \sum_{A, A'}  R_i^A(\mathbf{k}, \hat{l}_{ij}, \hat{l}_{ik} ) {R_l^{A'}}^*(\mathbf{k}, \hat{l}_{lm}, \hat{l}_{ln} )  \frac{ P_h^{A A' } (k) }{4 \pi k^2}  , 
\end{aligned}
\end{equation}
where we have used
\begin{equation}
\label{eq:spectrum}
 \langle \tilde{h}_{A_1} (\mathbf{k}_1) \tilde{h}^*_{A_2} (-\mathbf{k}_2)\rangle 
= \delta(\mathbf{k}_1 +\mathbf{k}_2) \frac{ P_h^{A_1 A_2} (k_1) }{4 \pi k_1^2} \; ,  \hspace{1cm} \langle \tilde{h}_{A_1} (\mathbf{k}_1) \tilde{h}_{A_2} (\mathbf{k}_2)\rangle 
 =\ 0 \; ,
\end{equation}
and we have taken $P_h^{A_1 A_2} (k_1)$ to be real. Moreover, since we restrict ourselves to a background with vanishing Stokes parameters $Q$ and $U$~\cite{Maggiore:2018sht}, $P_h^{A_1 A_2} (k_1)$ is diagonal in the $L/R$ basis:
\begin{equation}
P_h^{LL}(k) = I - V \;, \qquad P_h^{RR}(k) = I + V  \;.
\end{equation}
Therefore, $V$ must be vanishing too for the assumed, non-chiral background (\emph{i.e.} $P_h^{LL}(k) = P_h^{RR}(k)$). 

Finally~\cref{eq:two_point_dfreq} can be written as
\begin{equation}
\hspace{-1.2cm} \left\langle \Delta F_{i(jk)}(t) \Delta F_{l(mn)}(t)  \right\rangle  =  \sum_{A}  \int \textrm{d} k \;   2(2\pi k L)^2 |	W(kL) |^2 \delta_{AA'} \tilde{R}^{AA'}_{il(jk)(mn)}(k) \; P_h^{A } (k)  \; , 
\end{equation}
where
\begin{equation}
    \tilde{R}^{AA'}_{il(jk)(mn)}(k) \equiv \frac{1}{4 \pi }\int \textrm{d}^2 \hat{k} \;  \textrm{e}^{ -2\pi i  \mathbf{k} \cdot (\mathbf{x}_i - \mathbf{x}_l) }  R_i^A(\mathbf{k}, \hat{l}_{ij}, \hat{l}_{ik} ) R_l^{A' \ *}(\mathbf{k}, \hat{l}_{lm}, \hat{l}_{ln} ) \; .
\end{equation}
Since for LISA we have $ \tilde{R}^{LL}_{il(jk)(mn)} =  \tilde{R}^{RR}_{il(jk)(mn)} \equiv \tilde{R}_{il(jk)(mn)}$, for $V = 0$ we obtain
\begin{equation}
\label{eq:final_dfreq}
 \left\langle \Delta F_{i(jk)}(t) \Delta F_{l(mn)}(t)  \right\rangle = \int \textrm{d} k \;  4\,(2\pi k L)^2 |	W(kL) |^2  \tilde{R}_{il(jk)(mn)}(k) \; P_h (k)  \; , 
\end{equation}
with $ P_h (k) \equiv P_h^L(k) = P_h^R(k)$. At this point it is convenient to define the integrand (up to the factor $P_h(k)$) as
\begin{equation}
\label{eq:cal_R_def}
\mathcal{R}_{ij}(k) \equiv 4\, (2\pi k L)^2 |	W(kL) |^2  \tilde{R}_{il(jk)(mn)}(k)
\; .
\end{equation}
Notice that $\Delta F_{i(jk)}(t)$ is precisely the quantity labelled $s_i(t)$ in the main text.

\subsection{Noise spectra}
\label{sec:noise}
For completeness, we also include a review of the calculation of the noise spectra used in the main text. We assume that the laser frequency noise has been removed through time-delay interferometry. The dominant residual sources of noise are then associated with the interferometer measurement system (IMS), in particular shot noise for the light traveling between spacecrafts, and acceleration noise, fluctuations in the proof mass velocities, caused by residual electrostatic and magnetic forces on the proof masses. 

For concreteness, we assume that the laser beam is reflected off the proof mass upon arrival for transmission between optical benches on different spacecrafts, and before departure for transmission between optical benches on the same spacecraft. In this case, the acceleration noise contribution to $\tilde{d}_\text{X}$ is~\cite{Estabrook:2000ef, Tinto:2014lxa} 
\begin{eqnarray}
\hskip -1cm\tilde{d}^{\rm acc}_{\text{X} } &=& 
 \hphantom{+}\left(1+e^{4 \pi i  f L_{12}}\right) \left(e^{4 \pi i  f
   L_{13}}-1 \right) \hat{l}_{12}\mathbf{v}_{1L} + 2 e^{2 \pi i  f L_{13}} \left(1-e^{4 \pi i  f
   L_{12}}\right) \hat{l}_{31}\mathbf{v}_{3L}\nonumber\\ 
&& -\left(1+e^{4 \pi i  f L_{13}}\right) \left(e^{4 \pi i  f L_{12}}-1\right) \hat{l}_{13} \mathbf{v}_{1R}
 - 2 e^{2 \pi i  f L_{12}} \left(1 - e^{4 \pi i  f L_{13}}\right) \hat{l}_{21}\mathbf{v}_{2R}\,,
\end{eqnarray}
where $L_{ij}$ is the distance between spacecrafts $i$ and $j$, which is assumed to be time-independent, $\hat{l}_{ij}$ is a unit vector pointing from spacecraft $j$ to $i$ as before, and $\mathbf{v}_{iL}$ and $\mathbf{v}_{iR}$ are the fluctuations in the velocities of the proof mass in the left and right optical bench on spacecraft~$i$, respectively. The expressions for the variables $\tilde{d}^{\rm acc}_{\text{Y} }$ and $\tilde{d}^{\rm acc}_{\text{Z} }$ can be obtained by cyclic permutation.

If we assume that the fluctuations of the different probe masses are uncorrelated (even for masses on the same spacecraft) and are all characterized by the same power spectrum
\begin{equation}
\langle \hat{n}_{12}\tilde{\mathbf{v}}_{1L}(f) \hat{n}_{12}\tilde{\mathbf{v}}^*_{1L}(f')\rangle'=\langle \hat{n}_{13}\tilde{\mathbf{v}}_{1R}(f) \hat{n}_{13}\tilde{\mathbf{v}}^*_{1R}(f')\rangle' =\cdots= P_{\rm acc}(f)\,,
\end{equation}
the acceleration noise contribution to the power spectra are
\begin{eqnarray}
\hskip -1cm\langle \tilde{d}_\text{X}(f)\tilde{d}_\text{X}^*(f')\rangle'&=&8 \left(2 \sin ^2\left(2 \pi  f L_{12}\right)+2 \sin ^2\left(2 \pi  f L_{13}\right)\right.\nonumber\\
&&\hskip 0.3cm \left.+\sin ^2\left(2 \pi  f\left(L_{12}-L_{13}\right)\right)+\sin ^2\left(2 \pi  f
   \left(L_{12}+L_{13}\right)\right)\right)P_{\rm acc}(f)\,,\\
\hskip -1cm\langle \tilde{d}_\text{X}(f) \tilde{d}_\text{Y}^*(f')\rangle'&=&-32 e^{2 \pi i  f \left(L_{13}-L_{23}\right)} \sin \left(2 \pi  f
   L_{13}\right) \sin \left(2 \pi  f L_{23}\right) \cos \left(2 \pi  f
   L_{12}\right)P_{\rm acc}(f)\,.
\end{eqnarray}

We can proceed in the same way for IMS noise, and will assume it is dominated by shot noise. In this case, we can neglect the contribution from transmission between the optical benches on a single spacecraft, and will collect all sources of IMS noise for transmission from spacecraft $j$ to $i$ into a single term $\tilde{n}_{ij}$. The IMS noise contribution to $\tilde{d}_\text{X}$ then is
\begin{eqnarray}
\hskip -1cm\tilde{d}^{\rm IMS}_{\text{X} } &=& 
 \hphantom{+}\left(e^{4 \pi i  f L_{13}}-1\right) \tilde{n}_{12} + e^{2 \pi i f L_{13}}(1-e^{4 \pi i  f L_{12}})\tilde{n}_{31}\nonumber\\ 
&& -\left(e^{4 \pi i f L_{12}}-1\right) \tilde{n}_{13} 
 - e^{2 \pi i  f L_{12}}(1-e^{4\pi i  f L_{13}}) \tilde{n}_{21}\,.
\end{eqnarray}
The corresponding expressions for the variables $\tilde{d}^{\rm acc}_{\text{Y} }$ and $\tilde{d}^{\rm acc}_{\text{Z} }$ can again be found by cyclic permutation. 

If we assume that the IMS contributions for the different Doppler variables uncorrelated so that the noise covariance matrix is diagonal, and that the auto-spectra have identical statistical properties so that
\begin{equation}
\langle \tilde{n}_{12}(f)\tilde{n}_{12}(f')\rangle'=\langle \tilde{n}_{23}(f)\tilde{n}_{23}(f')\rangle'=\dots=P_{\rm IMS}(f)\,,
\end{equation}
the IMS noise contribution to the power spectra are
\begin{eqnarray}
\hskip -1cm\langle \tilde{d}_\text{X}(f)\tilde{d}_\text{X}^*(f')\rangle'&=&8\left( \sin ^2\left(2 \pi  f L_{12}\right)+ \sin ^2\left(2 \pi  f
   L_{13}\right)\right)P_{\rm IMS}(f)\,,\\
\hskip -1cm\langle \tilde{d}_\text{X}(f) \tilde{d}_\text{Y}^*(f')\rangle'&=&-8 e^{2 \pi i  f \left(L_{13}-L_{23}\right)} \sin \left(2 \pi  f L_{13}\right) \sin \left(2 \pi  f L_{23}\right) \cos \left(2 \pi  f L_{12}\right) P_{\rm IMS}(f)\,.
\end{eqnarray}
So far we have assumed that $L_{ij}$ are time-independent but not necessarily equal. The noise spectra in the main text assume an equilateral configuration, and can be obtained by setting $L_{12}=L_{13}=L_{23}=L$. Collecting both acceleration and IMS noise, we find
\begin{eqnarray}
\hskip -1cm\langle \tilde{d}_\text{X}(f)\tilde{d}_\text{X}^*(f')\rangle'&=&16\sin ^2\left(2 \pi  f L\right)\left\{\left(3+\cos\left(4 \pi  f L\right)\right)P_{\rm acc}(f)+P_{\rm IMS}(f)\right\}\,,\\
\hskip -1cm\langle \tilde{d}_\text{X}(f) \tilde{d}_\text{Y}^*(f')\rangle'&=&-8 \sin^2 \left(2 \pi  f L\right)  \cos \left(2 \pi  f L\right) \left[4P_{\rm acc}(f)+P_{\rm IMS}(f)\right]\,.
\end{eqnarray}
We see that for the equilateral configuration the noise covariance matrix for $\tilde{d}_\text{X}$, $\tilde{d}_\text{Y}$, and $\tilde{d}_\text{Z}$ becomes real. We also see that we can approximate the configuration as equilateral if $2\pi f \Delta L/c\ll1$.

\bibliographystyle{JHEP}
\bibliography{references.bib}

\end{document}